\documentclass[oneside,swedish,british,amsmath,amssymb,endnotes]{book}
\usepackage[T1]{fontenc}
\usepackage[latin9]{inputenc}
\usepackage{geometry}
\usepackage{color}
\usepackage{babel}
\usepackage{verbatim}
\usepackage{rotating}
\usepackage{units}
\usepackage{amsmath}
\usepackage{amssymb}
\usepackage{graphicx}
\usepackage{esint}
\usepackage[unicode=true,pdfusetitle,
 bookmarks=true,bookmarksnumbered=false,bookmarksopen=false,
 breaklinks=false,pdfborder={0 0 1},backref=false,colorlinks=false]
 {hyperref}

\makeatletter

\providecommand{\tabularnewline}{\\}

\usepackage[intoc]{nomencl}
\makenomenclature 
\usepackage{tikz}
\usepackage[perpage,symbol]{footmisc}

\makeatother

\begin{document}
\global\long\def\i{\imath}

\global\long\def\plll{\mathcal{P}_{\mathrm{LLL}}}

\global\long\def\plcs{\mathcal{P}_{\mathrm{LCS}}}

\global\long\def\translate#1{t\negmedspace\left(#1\right)}

\global\long\def\elliptic#1#2#3{\vartheta_{#1}\!\left(#2\middle|#3\right)}

\global\long\def\shortelliptic#1#2{\vartheta_{#1}\left(#2\right)}

\global\long\def\ellipticgeneralized#1#2#3#4{\vartheta\!\left[\begin{array}{c}
 #1\\
#2 
\end{array}\right]\!\left(#3\middle|#4\right)}

\global\long\def\ket#1{\left|#1\right\rangle }

\global\long\def\bra#1{\left\langle #1\right|}

\global\long\def\braket#1#2{\left\langle #1\left|\vphantom{#1}#2\right.\right\rangle }

\global\long\def\ketbra#1#2{\left|#1\vphantom{#2}\right\rangle \left\langle \vphantom{#1}#2\right|}

\global\long\def\braOket#1#2#3{\left\langle #1\left|\vphantom{#1#3}#2\right|#3\right\rangle }

\global\long\def\colred#1{\textcolor{red}{#1}}

\begin{titlepage}

\begin{minipage}[t]{1\columnwidth}%
\begin{center}
\vskip 1cm
\par\end{center}

\begin{center}
\includegraphics[scale=0.3]{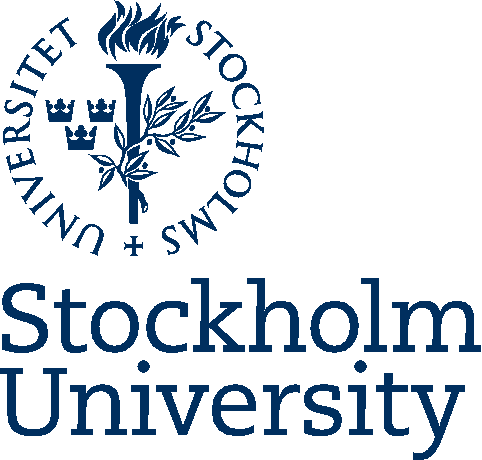}
\par\end{center}%
\end{minipage}

\vskip 1cm

\begin{minipage}[t]{1\columnwidth}%
\begin{center}
{\Huge Coherent State Wave Functions}
\par\end{center}%
\end{minipage}

\vskip .5cm

\begin{minipage}[t]{1\columnwidth}%
\begin{center}
{\Huge on the Torus}
\par\end{center}%
\end{minipage}

\vskip 1.5cm

\begin{minipage}[t]{1\columnwidth}%
\begin{center}
{\huge Mikael Fremling}
\par\end{center}%
\end{minipage}

\vskip 1cm

\begin{minipage}[t]{1\columnwidth}%
\begin{center}
Licentiate Thesis
\par\end{center}%
\end{minipage}

\vskip 1cm

\begin{minipage}[t]{1\columnwidth}%
\selectlanguage{swedish}%
\begin{center}
Akademisk avhandling\foreignlanguage{british}{ }\\
för\foreignlanguage{british}{ }avläggande\foreignlanguage{british}{
}av\foreignlanguage{british}{ }licentiatexamen i teoretisk fysik\foreignlanguage{british}{}\\
vid Stockholms Universitet\foreignlanguage{british}{ }
\par\end{center}\selectlanguage{british}%
\end{minipage}

\vskip 1cm

\begin{minipage}[t]{1\columnwidth}%
\begin{center}
Department of Physics\\
Stockholm University 
\par\end{center}%
\end{minipage}

\vskip 1cm

\begin{minipage}[t]{1\columnwidth}%
\begin{center}
{\Large Maj 2013}
\par\end{center}%
\end{minipage}\end{titlepage}

\newpage{}

\thispagestyle{empty}
\mbox{}

\newpage{}

\section*{\begin{center}\vskip 2cmAbstract\end{center}}

In the study of the quantum Hall effect there are still many unresolved
problems. One of these is how to generate representative wave functions
for ground states on other geometries than the planar and spherical.
We study one such geometry, the toroidal one, where the periodic boundary
conditions must be properly taken into account.

As a tool to study the torus we investigate the properties of various
types of localized states, similar to the\emph{ coherent states} of
the harmonic oscillator, which are maximally localized in phase space.
We consider two alternative definitions of localized states in the
lowest Landau level (LLL) on a torus. One is the projection of the
coordinate delta function onto the LLL. Another definition, proposed
by Haldane \& Rezayi, is to consider the set of functions which have
all their zeros at a single point. Since all LLL wave functions on
a torus, are uniquely defined by the position of their zeros, this
defines a set of functions that are expected to be localized around
the point maximally far away from the zeros. 

These two families of localized states have many properties in common
with the coherent states on the plane and on the sphere, \emph{e.g.}
a simple resolution of unity and a simple self-reproducing kernel.
However, we show that only the projected delta function is maximally
localized.

We find that because of modular covariance, there are severe restrictions
on which wave functions that are acceptable on the torus. As a result,
we can write down a trial wave function for the $\nu=\frac{2}{5}$
state, that respects the modular covariance, and has good numerical
overlap with the exact coulomb ground state.

Finally we present preliminary calculations of the antisymmetric component
of the viscosity tensor for the proposed, modular covariant, $\nu=\frac{2}{5}$
state, and find that it is in agreement with theoretical predictions.

\pagebreak{}

\section*{\vskip 2cmAcknowledgements}

I would like to thank my two supervisors Hans Hansson and Anders Karlhede
for support and inspiration. It must be frustrating when minor bugs
change the result from success to failure and back again. Thank you
all friends and colleagues who in one way or another have contributed
to this thesis, whether it be proofreading, being bugged with questions
or just general discussions. A special thanks goes to Gertrud Fremling
for thoroughly proofreading the manuscript, I do not want to think
of what it would have looked like if you had not. I would also like
to thank my wife, Karin Fremling, who has not only put up with my
frequent absentmindedness, but also encouraged my work wholeheartedly.

Finally, I would like to thank YOU, the reader of this thesis, for
showing an interest in my work.\\
\\
Thank you!

\pagebreak{}

\tableofcontents{}

\printnomenclature

\chapter*{List of accompanying papers}

\addcontentsline{toc}{chapter}{List of Acompanying Papers}

\begin{tabular}{ll}
{\large Paper I} & \textbf{\large Coherent State Wave Functions on a }\tabularnewline
 & \textbf{\large Torus with a Constant Magnetic Field}\tabularnewline
 & {\large M. Fremling}\tabularnewline
 & {\large J. Phys. A, under consideration {[}arXiv:1302.6471{]} (2013)}\tabularnewline
 & \tabularnewline
{\large Paper II} & \textbf{\large Hall viscosity of hierarchical}\tabularnewline
 & \textbf{\large quantum hall states.}\tabularnewline
 & {\large M. Fremling, T. H. Hansson, and J. Suorsa.}\tabularnewline
 & {\large In preparation, (2013)}\tabularnewline
\end{tabular}

\chapter{Introduction and Outline }

This year marks the 30 year anniversary of Laughlin's famous $\nu=\frac{1}{3}$
wave function, introduced to explain the Fractional Quantum Hall Effect.
With the Laughlin wave function came the notion of excitations with
fractional charge, and fractional statistics. The theory of the Quantum
Hall Effect is still an active area of research. The Integer Hall
Effect was the first example of a Topological Insulator\cite{Kane_2005},
but many others have been proposed and realized. Fractional charges
have also been proposed to exist in other types of systems, where
fractional Chern Insulators are a case in point\cite{Regnault_11}.
Vivid research has also been focused on the special state at $\nu=\frac{5}{2}$.
This state is expected to support excitations with non-abelian braiding
properties. The non-abelian statistics makes this state of matter
an interesting candidate for quantum information storage and processing;
in short, a quantum computer.

In quantum mechanics, the existence of a magnetic field drastically
alters the structure of the Hilbert space as compared to the case
of free particles. The continuum of energy levels of the free particle,
transforms into highly degenerate Landau levels with a degeneracy
proportional to the strength of the magnetic field. If the applied
magnetic field is strong enough, together with low temperatures, and
clean samples, the Quantum Hall Effect is observed. The Fractional
Quantum Hall Effect (FQHE) is observed in high quality semiconductor
junctions, but also in graphene. In semiconductors the temperature
has to be low for the FQHE to be manifested, but in graphene the effect
is observable even at room temperature\cite{Novoselov_2005}.

Both the Integer and the Fractional Quantum Hall Effects are examples
of Topological Insulators; States of matter that are insulating in
the bulk, but has dissipationless transport at the edges. The topological
aspect of the FQHE is that it is insensitive to continuous deformations
of the geometry of a sample, but also to small variations of the applied
magnetic field, or temperature. Most importantly, the dissipationless
edge currents even survive a finite amount of impurities, which is
always present in a real system. A consequence of this is that the
electric resistance $R_{H}$ is quantized, to an experimentally very
high accuracy.

The topology of a state is important, and not all probes can detect
topological quantities. Especially local measurement should not be
able to distinguish between a topological and a trivial insulator. 

In this thesis we are studying the FQHE on the torus. This is interesting
as one of the topological aspect is encoded in the ground state degeneracy
on the torus. The torus is also a good playground to test model trial
functions coming from Conformal Field Theory (CFT). Trial wave functions
for the FQHE have been deduced using correlators from CFT. The CFT
wave functions are easily evaluated in a planar geometry, but numerical
comparison to exact coulomb ground states can be difficult to perform
because of boundary effects. The torus represents a natural arena
to for numerical tests.

When constructing FQH-wave functions, the CFT trial wave functions
need to be projected to the lowest Landau Level, to obtain physical
electronic wave functions. The projector to the lowest Landau Level
can naturally be expressed of in terms of coherent states. For that
reason a more careful study of coherent states on a toroidal geometry
is needed. In this thesis we study the basic properties of coherent
states on a torus. We consider study two kinds of coherent states,
and their various properties.

In addition to studying coherent states on a torus we also investigate
how to generate trial wave functions on the torus, in a self-consistent
manner. As a result we find that modular properties strongly constrain
the possible wave functions on the torus, and we propose a trial wave
function for the $\nu=\frac{2}{5}$ state that has the correct modular
properties.

Using the proposed wave function, we calculate a topological characteristic
of the quantum Hall system; the antisymmetric component of the viscosity
tensor. Read has demonstrated that the viscosity is proportional to
the mean orbital spin of the electron, which is a topological quantity.
This transport coefficient can be measured numerically by changing
the geometry of the torus\cite{Read_11}.

This thesis has two accompanying papers. The first is my own work
on coherent states, and the second, in preparation, is in collaboration
with my supervisor Thors Hans Hansson, and Juha Suorsa at Nordita.

\pagebreak{}

\chapter{The Quantum Hall Effect}

\section{The Classical Hall Effect}

In 1879 the American physicist Edwin Hall decided to test whether
or not electric currents where affected by magnetic forces\cite{Hall_1879}.
He designed an experiment in which he found that a thin metal plate
in a magnetic field \textbf{$\mathbf{B}$,} perpendicular\textbf{
}to the surface of the plate, will experience a voltage drop in a
direction perpendicular to $\mathbf{B}$ and the current $\mathbf{I}$
flowing through the plate. He concluded that the perpendicular resistance
$R_{\mathrm{H}}=\frac{V_{\perp}}{I}$\nomenclature[2]{$R_H$}{Hall resistance}
was proportional to the strength of the magnetic field and sensitive
to the sign of the magnetic field.

The Hall Effect is explained by the behaviour of charged particles
in a magnetic field. As the electrons move though the magnetic field,
they will be subject to a Lorenz force $\mathbf{F}_{B}=q\mathbf{v}\times\mathbf{B}$
directed toward one of the edges of the plate. As more and more electrons
are diverted toward one side, a charge imbalance built up inside the
plate generating an electric field across the plate. The existence
of a static electric field means that there a voltage difference,
which in this case will be perpendicular to the direction of the current
$\mathbf{I}$. Eventually the electric field, with the associated
electric force $\mathbf{F}_{E}=q\mathbf{E}$, will be large enough
to balance the magnetic force $\mathbf{F}_{B}$. This voltage drop
must be proportional to the total current, as a larger current increases
the number of electrons that are being diverted. The voltage difference
must also be proportional to the magnetic field, as the Lorenz force
that deflects electrons is proportional in strength to $B$. Hence,
the Hall resistance, which is the perpendicular resistance $R_{H}$,
is proportional to the strength of magnetic field $R_{\mathrm{H}}\propto B$.
The Hall Effect is also inversely proportional to the thickness of
the material the current runs through, which means that the Hall Effect
gets stronger the thinner the plate is. A more careful analysis shows
that the Hall Resistance is $R_{H}=\frac{B}{e\rho_{3D}d}$, where
$d$ is the thickness of the plate, and $\rho_{3D}$ is the electron
density. In the limit of very thin plates, that are almost two dimensional,
$R_{H}$ is better described using the the two dimensional density
$\rho_{2D}$, as $R_{H}=\frac{B}{e\rho_{2D}}$. It is in this limit
of thin plates that quantum mechanical effects can become important,
and the Hall Effect can be changed into the Quantum Hall Effect.

\begin{figure}
\begin{centering}
\includegraphics[width=0.6\columnwidth]{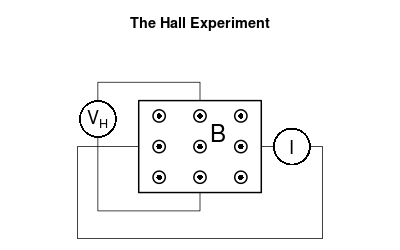}
\par\end{centering}

\caption{The Hall experiment. A current $I$ is driven through a thin metal
plate with a perpendicular magnetic field $B$ such that a voltage
$V$ is measured in the transverse direction.\label{fig:Hall_experiment}}
\end{figure}

\section{The Quantum Hall Effect}

In 1980 von Klitzing gave the Hall Effect a new twist\cite{Klitzing_80}
by confining electrons to two dimensions, in semiconductor junctions.
In his experiments, where he had high quality samples in combination
with low temperatures and high magnetic fields, the Hall resistance
$R_{H}$ deviated from the classically predicted linear behaviour
and instead started developing kinks and plateaus. Furthermore, these
plateaus appeared at regular intervals such that the resistance at
the plateaus were given by the formula $R_{H}=\frac{1}{\nu}\cdot\frac{h}{e^{2}}$,
where $\nu$ is an integer. In addition, at the magnetic fields where
the plateaus appeared in the Hall resistance, the longitudinal resistance
$\mathbf{R}_{\parallel}$ dropped to zero. This new phenomena was
dubbed the Integer Quantum Hall Effect (IQHE)\nomenclature[1]{IQHE}{Integer Quantum Hall Effect}.
The IQHE is that precise that it effectively defines the unit of resistance.
The fundamental unit of resistance can be measured with an accuracy
of $10^{-12}$ to be $R_{K}=\frac{h}{e^{2}}=25812.807557(18)$ $\Omega$\cite{Tzalenchuk_10}.

\begin{figure}
\begin{centering}
\includegraphics[width=0.5\columnwidth]{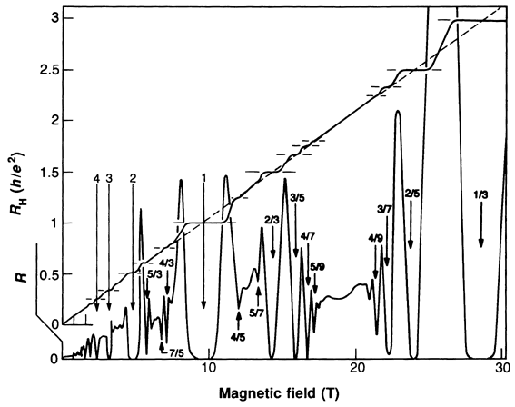}
\par\end{centering}

\caption{Resistance measurements of the FQHE. The transverse resistivity $R_{H}$
shows kinks and plateaus at $\nu=\frac{p}{q}$. At the same rational
fractions the longitudinal resistance $R$ drops to zero\cite{Stormer_92}.\label{fig:FQHE}}
\end{figure}

As samples became cleaner, and temperatures lower, new features appeared
in the resistance spectrum. New plateaus were observed, together with
dips in the longitudinal resistivity. These new plateaus where located
at $R_{H}=\frac{1}{\nu}\cdot\frac{h}{e^{2}}$, where $v=\frac{p}{q}$\nomenclature[2]{$\nu$}{Filling fraction}
\nomenclature[2]{$N_e$}{Number of Electrons} \nomenclature[2]{$N_\Phi$}{Number of Magnetic Flux Quanta}
\nomenclature[2]{$N_s$}{Number of states in the Hilbert space. On the torus $N_s=N_\Phi$, and $L_xL_y=2\pi N_s$}
formed fractions, such as $\frac{1}{3}$, $\frac{2}{5}$ and $\frac{3}{7}$.
The plateaus only developed at fractions with an odd denominator,
as can be seen in Figure \ref{fig:FQHE}. The new effect was named
Fractional Quantum Hall Effect (FQHE)\nomenclature[1]{FQHE}{Fractional Quantum Hall Effect}.
Compared to the IQHE it has more features beyond simply a fractional
Hall resistance. One prominent feature is that the minimal excitations
do not consist of individual electrons but rather of fractionally
charged quasi-particles\cite{Laughlin_83}, that do not obey the ordinary
statistics of fermions or bosons. This new form of statistics constitutes
a generalization of the fermion/boson statistics and can only be obtained
in systems with lower dimensionality than 3. Some of these quasi-particles
even display non-abelian statistics\cite{Moore_91}, in theory. The
experimental verification of the non-abelian statistics is still lacking,
but this is the reason that people are looking to FQHE as a means
of building a quantum computer.

The key to understanding the IQHE lies in the behaviour of single
particles in a magnetic field. From classical physics we know that
charged particles are deflected by magnetic fields and therefore move
in circles where the radius is proportional to the particle's momentum.
The frequency of revolution is therefore independent of the particle
momentum. It depends only on the magnetic field $\mathbf{B}$ and
on the mass $m$ of the particle, as expressed by the formula $\omega_{c}=\frac{eB}{mc}$.
The oscillatory behaviour is is similar to the behaviour of the Harmonic
Oscillator, where the quantum mechanical energy levels are equally
spaced as $E_{n}=\hbar\omega\left(n+\frac{1}{2}\right)$ with $n$
being an integer. An analogous calculation for a particle in a magnetic
field shows that here, too, the energy levels are equally spaced,
with $E_{n}=\hbar\omega_{c}\left(n+\frac{1}{2}\right)$. Each energy
level is called a Landau Level (LL)\nomenclature[1]{LL}{Landau Level},
after Landau\cite{Landau_30} who solved the problem in 1930. The
LL with $n=0$ is the minimum energy level and therefore called the
Lowest Landau Level (LLL)\nomenclature[1]{LLL}{Lowest Landau Level}.
In contrast to the Harmonic Oscillator, each LL is massively degenerate,
as there exists one state for each flux quanta $\Phi_{0}=\frac{h}{e}$
of the magnetic field. Thus the density of states in any Landau Level
is $\frac{B}{\Phi_{o}}\approx\frac{B}{\mbox{1 Tesla}}\times242$ per
$\left(\mu\mathrm{m}\right)^{2}$. This means that if each electron
were confined to a circle, the radius of that circle would be $r=\sqrt{\frac{\Phi_{o}}{\pi B}}=\unit[363]{\mathring{A}}\times\sqrt{\frac{\mbox{1 Tesla}}{B}}$.
It is customary to introduce a length scale $\ell=\frac{r}{\sqrt{2}}$
known as the magnetic length\nomenclature[2]{$\ell$}{Magnetic length: $\ell=\sqrt{\frac h {eB}}$}.

The above mentioned factor $\nu$ can be calculated as the filling
factor $\nu=\frac{N_{e}}{N_{s}}$, which counts the number of filled
Landau levels. If $\nu$ is an integer, all the Landau levels up to
level $\nu$ are completely filled. Thus there exists a gap of $\hbar\frac{eB}{mc}$
to excite an electron into the next LL\cite{Laughlin_81}. This gap
causes the IQH-state to be stable against small variations in the
magnetic field, as the energy cost of moving an electron to the next
LL would be too large.

For the FQHE the explanation is not as straight forward as for the
IQHE. As $\nu$ is no longer an integer, but rather a fraction, such
as $\nu=\frac{1}{3}$. One LL will be only partially filled, so the
single particle picture of electrons filling one or more entire LLs
no longer works. In order to solve this problem we need to go beyond
the properties of individual electrons. The answer lies in studying
the interaction between the particles within a LL. Crudely speaking,
the Coulomb repulsion between electrons forces any two electrons to
be as far separated in space as possible. This results in a highly
correlated fluid where the minimal excitation has a finite energy.

Both the IQHE and the FQHE needs some amount of impurities to manifest
themselves. If the sample would be fully translationally invariant,
then Lorentz invariance would imply that no plateaus can be present.
Impurities are needed to break the Lorentz invariance. However, if
the impurities are too strong, then the QHE is not observable. Herein
lies the reason why not all FQHE fractions are observable in experiments.
In the limit of no impurities, all FQHE fractions will be visible,
but this will result in a devil's staircase of plateaus in $R_{H}$.
In that case, FQHE becomes indistinguishable from the classical Hall
Effect, at least in simple transport experiments.

\section{The Laughlin Construction and the Hierarchy\label{sub:Laughlin-Hierarchy}}

In 1983 Robert Laughlin proposed a wave function that would explain
the FQHE at $\nu=\frac{1}{q}$, where $q$ in an odd integer\cite{Laughlin_83}.
The construction was inspired by the realization that in the FQH-states
the electrons could minimize their interaction energy by being as
far from each other as possible. With that as a guiding star, he proposed
the now famous wave function 
\begin{equation}
\Psi_{\frac{1}{q}}\left(z_{1},\dots,z_{N_{e}}\right)=e^{-\frac{1}{4}\sum_{j}\left|z_{j}\right|^{2}}\prod_{i<j}^{N_{e}}\left(z_{i}-z_{j}\right)^{q},\label{eq:Lauhlin-wfn}
\end{equation}
 which is a homogeneous state with well-defined angular momentum.
This wave function implied that only odd denominator filling fractions
could appear, since otherwise the wave function would not be antisymmetric
in the electron coordinates. Starting from \eqref{eq:Lauhlin-wfn}
he could also find the elementary excitations, also called quasi-particles,
that could appear. This was accomplished by inserting an extra quantum
of flux into the state at $z=\eta$, and noting that the new wave
function contained an extra factor $\prod_{j}\left(z_{j}-\eta\right)$.
By making an analogy with a charged plasma, Laughlin could deduce
that the quasi-particles at $\nu=\frac{1}{q}$ have fractional charges
$\frac{e}{q}$. The physical picture is that the term $\left(z-\eta\right)$
does not repel the electron and quasi-paricle as strongly as the $\left(z_{i}-z_{j}\right)^{q}$
repels the electrons from each other. This gives the quasi-particle
a smaller correlation hole that the electron. Later Arovas, Schrieffer
and Wilczek deduced that the quasi-particles have fractional exchange
statistics\cite{Arovas_84}.

The Laughlin wave function sheds some light on other filling fractions
as well, since the quasi-particle excitations can be used as building
blocks for other states. As the magnetic field $B$ is tuned away
from $\nu=\frac{1}{q}$, quasi-particles appear in the state \eqref{eq:Lauhlin-wfn}.
As $B$ is tuned still further, these quasi-particles becomes so numerous
that the electrons and quasi-particles condense into a new state,
with a new filling fraction. This new state will also support its
own quasi-particles with fractional charges and statistics. These
$2^{\mathrm{nd}}$ level quasi-particles can in turn, as the magnetic
field is changed further, condense into yet another state. By this
process any filling fraction with an odd denominator can be created
by continuous condensation of parent quasi-particles\cite{Haldane_83,Halperin_83}.
This idea is called the Haldane-Halperin hierarchy construction, since
different filling fractions come at different hierarchical levels
of condensation of quasi-particles.

Each level of the hierarchy have both negatively and positively charged
quasi-particles. The negatively%
\footnote{Negative charge with respect to the electron charge.%
} charged excitations are called quasi-holes. Depending on if quasi-particles
or quasi-holes are condensated, different technical issues arise.
Usually quasi-particle condensation is more simple and quasi-hole
condensation more difficult.

In the hierarchy, all quasi-particle excitations are gapped, compared
to the ground state. This gap sheds some light in which order the
different fractions should become visible in experiments. If the FQHE
is to be measured, it is important that the gap to quasi-particle
excitation is not bridged by temperature or impurities. It can be
shown, under certain circumstances, that the excitation gap of the
FQHE at $\nu=\frac{p}{q}$ is monotonically vanishing in $q$\cite{Bergholtz_08}.
This explains why the fractions at $\nu=\frac{1}{3}$ and $\nu=\frac{2}{3}$
are observed first, followed by the fractional at $\nu=\frac{2}{5}$,
$\nu=\frac{3}{7}$ and $\nu=\frac{4}{9}$ etc.

\section{Composite Fermions}

A different route to explaining the FQHE was taken by Jain. Inspired
by Laughlin's wave function and resistance measurements, he unified
the FQHE and the IQHE by introducing the notion of composite fermions\cite{Jain_07_Book}.
Jain proposed that the electrons could screen parts of the magnetic
field by binding vortices to themselves. By binding just enough vortices,
which reduces the magnetic field, the electrons would fill one or
more effective LLs. This construction yielded explicit expressions
for wave functions at other filling fractions than $\nu=\frac{1}{q}$,
something that the hierarchy construction could not achieve. Furthermore,
Jain found that the wave functions for Composite Fermions also displayed
remarkably good overlap, with those obtained from exact diagonalization
of the Coulomb potential.

There now exists an alternative method for deducing trial wave functions
for generic FQH-states, based on the similarity between the Laughlin
wave function and correlators in Conformal Field Theory (CFT) \nomenclature[1]{CFT}{Conformal Field Theory}.
These CFT-based wave functions, reproduce the wave functions deduced
using the composite fermion picture. Thus the composite fermion scheme
can be seen as a special case of the hierarchy construction and implies
that these two approaches are two alternative ways of looking at the
same problem.

\section{Fractional Quantum Hall Effect on a Torus}

In this licentiate thesis we will consider the Haldane-Halperin hierarchy
wave functions in a toroidal geometry. By construction, the torus
lacks a boundary, making it suitable for numerical calculations. The
torus is also locally flat, which avoids the trouble that is connected
to the curved space of the sphere -- another geometry that lacks boundaries.
Further, the number of states in the torus Hilbert space is the same
as the number of magnetic flux quanta $N_{s}=\frac{A}{2\pi\ell^{2}}$,
where $A$ is the torus area.

The torus does of course come with its own set of problems. Because
of the periodicity, wave functions expressed on the torus have rather
complicated analytical forms. This includes products of Jacobi $\vartheta$-functions
$\elliptic jz{\tau}$, making analytical manipulations more complicated.
Also because of the gauge field associated with the magnetic field,
the wave functions are not truly periodic, as there is a restriction
on which translation operators that are allowed on the torus.

Examining this restriction will form a central part of this thesis.
This is an interesting problem, as this restriction prohibits the
mapping of CFT wave functions formulated on the plane directly to
the torus. Technically this is because the planar wave functions in
the higher levels of the Haldane-Halperin hierarchy will contain derivative
operators $\partial_{z}$. We will later show that these derivatives
can \emph{not} be interpreted as derivatives on the torus. Instead
the derivative can, at best, be mapped onto a linear combination of
allowed translation operators $t_{x}$ as $\partial_{z}\rightarrow\sum_{l}a_{l}t_{x}^{l}$.
The precise meaning of derivatives and translation operators will
be clarified in Section \ref{sec:The-torus} and \ref{sec:Derivatives}.

\pagebreak{}

\chapter{Coherent States in a Magnetic Field}

Coherent states can be thought of as the quantum mechanical analogue
of classical states. There are several ways of defining coherent states,
but in the simplest cases they are maximally localized in phase space.
The coherent states also obey the classical equations of motion. 

In order to set the stage for coherent states on torus, we will review
the concept of coherent states in general. As a warm-up, and to set
the notation, we will construct the coherent states in the Harmonic
Oscillator. We will then construct coherent states in a magnetic field
on the plane. After that we will explain why the torus poses a problem
and why the methods we employed, for the Harmonic Oscillator and on
the plane, cannot be directly applied to the torus. Finally we will
then construct two candidates for coherent states on the torus and
analyse their properties.

We will in several sections characterize the states with the use of
the Heisenberg uncertainty relation. We therefore review its general
form and basic properties. The general form of the uncertainty relations
states that 
\begin{equation}
\sigma_{A}\sigma_{B}\geq\frac{1}{2}\left|\left\langle \left[A,B\right]\right\rangle \right|.\label{eq:Uncertianty-relation}
\end{equation}
We define the uncertainty $\sigma_{A}$ of an operator $A$ as $\sigma_{A}^{2}=\left\langle A^{2}\right\rangle -\left\langle A\right\rangle ^{2}$\nomenclature[2]{$\sigma_A$}{Standard deviation of the expectation value of the operator $A$},
where $\left\langle \mathcal{O}\right\rangle $ is an expectation
value with respect to the operator $\mathcal{O}$ for a specific state.
In the special case where of $\hat{x}$ and $\hat{p}$ the relation
\eqref{eq:Uncertianty-relation}reduces to $\sigma_{x}\sigma_{p}\geq\frac{1}{2}\hbar$
since $\left[x,p\right]=\i\hbar$ is just a complex number.

\section{Coherent States in the Harmonic Oscillator\label{sub:CS-in-HO}}

We begin by reviewing the coherent states in the Harmonic Oscillator.
The one-dimensional quantum Harmonic Oscillator has a Hamiltonian
$H=\frac{1}{2m}\hat{p}^{2}+m\omega^{2}\hat{x}^{2}$. Using a suitable
choice of variables, we may rewrite this $H$ as $H=\hbar\omega\left(a^{\dagger}a+\frac{1}{2}\right)$,
where 
\begin{eqnarray}
a & = & \sqrt{\frac{m\omega}{2\hbar}}\left(\hat{x}+\frac{\imath}{m\omega}\hat{p}\right)\label{eq:a_HO}\\
a^{\dagger} & = & \sqrt{\frac{m\omega}{2\hbar}}\left(\hat{x}-\frac{\imath}{m\omega}\hat{p}\right)\label{eq:a_dagger_HO}
\end{eqnarray}
 and $\left[a,a^{\dagger}\right]=1$. A complete basis of solutions
is given by the states that are eigenstates of $a^{\dagger}a$, such
that $a^{\dagger}a\ket n=n\ket n$. We seek states that fulfil the
equality in Heisenberg's uncertainty relation $\sigma_{x}\sigma_{p}\geq\frac{\hbar}{2}$,
and start by examining the states $\ket n$. For this calculation,
$\hat{x}$ and $\hat{p}$ are expressed in terms of $a$ and $a^{\dagger}$
as 
\begin{eqnarray}
\hat{x} & = & \sqrt{\frac{\hbar}{2m\omega}}\left(a^{\dagger}+a\right)\label{eq:HO_x_of_ladder}\\
\hat{p} & = & \imath\sqrt{\frac{m\omega\hbar}{2}}\left(a^{\dagger}-a\right).\label{eq:HO_p_of_ladder}
\end{eqnarray}
 For the state $\ket n$, it is straightforward to verify that $\braOket n{\hat{x}}n=\braOket n{\hat{p}}n=0$.
It is also simple to show that $\braOket n{\hat{x}^{2}}n=\frac{\hbar}{m\omega}\left(n+\frac{1}{2}\right)$
and that $\braOket n{\hat{p}^{2}}n=m\omega\hbar\left(n+\frac{1}{2}\right)$.
Putting all the pieces together the result is $\sigma_{x}\sigma_{p}=\hbar\left(n+\frac{1}{2}\right)$.
It is only the state $\ket 0$ that equates the uncertainty relation,
and this happens to be an eigenstate of the $a$ operator with eigenvalue
$0$. We may thus instead look for the eigenstates of $a$ and $a^{\dagger}$.
It is simple to verify that there are no eigenstates of $a^{\dagger}$.
The class of states that are eigenstates of $a$ are characterized
by a complex number $\alpha$ such that $a\ket{\alpha}=\alpha\ket{\alpha}$
and $\bra{\alpha}a^{\dagger}=\bra{\alpha}\alpha$. These normalized
states are the Coherent States (CS)\nomenclature[1]{CS}{Coherent States}
\begin{equation}
\ket{\alpha}=e^{-\frac{1}{2}\left|\alpha\right|^{2}}e^{\alpha a^{\dagger}}\ket 0=e^{\alpha a^{\dagger}+\alpha^{\star}a}\ket 0.\label{eq:HO_CS}
\end{equation}
 The states $\ket{\alpha}$ are not energy eigenstates but are instead
maximally localized in phase space. From \eqref{eq:HO_x_of_ladder}
and \eqref{eq:HO_p_of_ladder} it is easy to see that the state $\ket{\alpha}$
has $\left\langle x\right\rangle =\sqrt{\frac{2\hbar}{m\omega}}\Re\left(\alpha\right)$
and $\left\langle p\right\rangle =\sqrt{2m\omega\hbar}\Im\left(\alpha\right)$
as well as $\left\langle x^{2}\right\rangle =\frac{2\hbar}{m\omega}\left[\Re\left(\alpha\right)^{2}+\frac{1}{4}\right]$
and $\left\langle p^{2}\right\rangle =2m\omega\hbar\left[\Im\left(\alpha\right)^{2}+\frac{1}{4}\right]$.
This shows that these states indeed minimize $\sigma_{x}\sigma_{p}$
since the variance is $\sigma_{x}^{2}=\left\langle x^{2}\right\rangle -\left\langle x\right\rangle ^{2}=\frac{\hbar}{2m\omega}$
and $\sigma_{p}^{2}=\frac{1}{2}m\omega\hbar$ which gives the product
$\sigma_{x}\sigma_{p}=\frac{\hbar}{2}$. Note that $\alpha=0$ corresponds
to the ground state $\ket 0$ which is of course annihilated by $a$.

The states $\ket{\alpha}$ do not only saturate the Heisenberg uncertainty
relations, they also posses a time evolution that mimics that of a
classical particle. We know from the commutation relations that $\dot{\left\langle x\right\rangle }=\frac{1}{m}\left\langle p\right\rangle $
and $\dot{\left\langle p\right\rangle }=-m\omega^{2}\left\langle x\right\rangle $
such that the time evolution of $\ket{\alpha}$ is $\ket{\alpha_{0}e^{\imath\omega t+\i\phi}}$
with energy $\left\langle E\right\rangle _{\alpha}=\hbar\omega\left(\left|\alpha_{0}\right|^{2}+\frac{1}{2}\right)$.
These states are therefore moving on circles in phase space with expectation
value $\left\langle x\right\rangle =x_{\mathrm{max}}\cos\left(\omega t+\phi\right)$
where $x_{\mathrm{max}}=\sqrt{\frac{2\hbar}{m\omega}}\left|\alpha_{0}\right|$.
As these states are \emph{not} energy eigenstates, the uncertainty
in energy $\sigma_{E}=\sqrt{\left\langle E^{2}\right\rangle _{\alpha}-\left\langle E\right\rangle _{\alpha}^{2}}$
is finite, and equal to $\sigma_{E}=\hbar\omega\left|\alpha_{0}\right|$.

\section[Coherent States in a Magnetic Field in Planar Geometry]{Coherent States in a Magnetic Field in \protect \\
Planar Geometry\label{sub:CS-in-B-on-plane}}

In the previous section we saw that we could construct coherent states
in the Harmonic Oscillator as eigenstates of the ladder operators.
On a plane in a magnetic field, a similar thing happens, but with
two operators instead of one. The Hamiltonian for a particle in a
magnetic field is given by 

\begin{equation}
\hat{H}=\frac{1}{2m}\left(p_{y}-eA_{y}\right)^{2}+\frac{1}{2m}\left(p_{x}-eA_{x}\right)^{2}\label{eq:General hamiltonian}
\end{equation}
 where $\mathbf{A}=\left(A_{x,}A_{y},A_{z}\right)$ is a vector potential
such that $\mathbf{B}=\nabla\times\mathbf{A}$. Depending on the choice
of gauge, we may introduce suitable ladder operators such that the
Hamiltonian can again be written as $\hat{H}=\hbar\omega\left(a^{\dagger}a+\frac{1}{2}\right)$.
Here there are two dimensions, $x$ and $y$, so we may now construct
two kinds of ladder operators instead of one. One set of operators
are $a$ and $a^{\dagger}$, which step up and down in what we call
Landau levels. These operators change the energy of the state, just
like the ladder operators in the Harmonic Oscillator. The other set
of operators is $b$ and $b^{\dagger}$ which, in symmetric gauge,
change the angular momentum of the electron. These operators keep
the electrons within a given Landau level%
\footnote{Since neither $b$ nor $b^{\dagger}$ appear in the Hamiltonian these
operators map out a degenerate subspace in each Landau level.%
} and are thus responsible for the large degeneracy within each LL.
The operators have the usual ladder operator commutation relations
$\left[a,a^{\dagger}\right]=\left[b,b^{\dagger}\right]=1$ and $\left[a,b\right]=\left[a,b^{\dagger}\right]=0$.
Using these, we may construct the eigenstates of $a$ and $b$ such
that $a\ket{\alpha,\beta}=\alpha\ket{\alpha,\beta}$ and $b\ket{\alpha,\beta}=\beta\ket{\alpha,\beta}$.
In analogy with the Harmonic Oscillator, these states can be expressed
as 
\begin{equation}
\ket{\alpha,\beta}=e^{-\frac{1}{4}\left|\alpha\right|^{2}-\frac{1}{4}\left|\beta\right|^{2}}e^{\frac{1}{\sqrt{2}}\left(\alpha a^{\dagger}+\beta b^{\dagger}\right)}\ket 0\label{eq:CS_on_plane}
\end{equation}
 where an extra factor of $\frac{1}{\sqrt{2}}$ has been introduced
for later convenience. The state $\ket 0$ is destroyed by both $a$
and $b$. The Hamiltonian can be written as $\hat{H}=\hbar\omega\left(a^{\dagger}a+\frac{1}{2}\right)$
which means that $\alpha$ must be related to the orbital motion of
the electron whereas $\beta$ must be related to the guiding centre
of the motion.

Let us quantify this. In symmetric gauge, $\mathbf{A}=\frac{1}{2}B\left(y\hat{\mathbf{x}}-x\hat{\mathbf{y}}\right)$,
the ladder operators are given as 

\begin{eqnarray*}
a=\frac{1}{\sqrt{2}}\left(\frac{\bar{z}}{2}+2\partial_{z}\right) &  & b=\frac{1}{\sqrt{2}}\left(\frac{z}{2}+2\partial_{\bar{z}}\right)\\
a^{\dagger}=\frac{1}{\sqrt{2}}\left(\frac{z}{2}-2\partial_{\bar{z}}\right) &  & b^{\dagger}=\frac{1}{\sqrt{2}}\left(\frac{\bar{z}}{2}-2\partial_{z}\right)
\end{eqnarray*}
 where all the dimensional factors have been suppressed since we set
$\hbar=\omega=m=1$. We have also introduced complex coordinates as
$z=x+\i y$. Inverting the relations above means that the coordinate
and momentum operators can be expressed in terms of $a$ and $b$
as

\begin{eqnarray*}
z=\sqrt{2}\left(b+a^{\dagger}\right) &  & \partial_{z}=\frac{1}{2\sqrt{2}}\left(a-b^{\dagger}\right)\\
\bar{z}=\sqrt{2}\left(a+b^{\dagger}\right) &  & \partial_{\bar{z}}=\frac{1}{2\sqrt{2}}\left(b-a^{\dagger}\right).
\end{eqnarray*}
 We immediately see that the positions expectation value for the coherent
states is $\left\langle z\right\rangle =\left(\beta+\bar{\alpha}\right)$.
Calculating the time evolution of $\left\langle z\right\rangle $,
we get $\dot{\left\langle z\right\rangle }=\left\langle \frac{-\i}{\hbar}\left[z,H\right]\right\rangle =\i\omega\bar{\alpha}$
giving the solution $\left\langle z\right\rangle =\left(\beta+\alpha_{0}e^{\imath\omega t+\i\phi_{0}}\right)$.
We may interpret this as the electron circulating at a radius $\alpha_{0}$
around a guiding centre $\beta_{0}$ with a frequency of $\omega$.
This state has energy $\left\langle E\right\rangle _{\alpha,\beta}=H=\hbar\omega\left(\frac{1}{2}\left|\alpha_{0}\right|^{2}+\frac{1}{2}\right)$.

A different way of looking at \eqref{eq:CS_on_plane} is to first
create a coherent excitation centred at $z=0$ using $e^{-\frac{1}{4}\left|\alpha\right|^{2}}e^{\frac{1}{\sqrt{2}}\alpha a^{\dagger}}$,
and then move to $z=\beta$ by $e^{-\frac{1}{4}\left|\beta\right|^{2}}e^{\frac{1}{\sqrt{2}}\beta b^{\dagger}}$.
The operator $e^{\beta b^{\dagger}-\bar{\beta}b}$ can be interpreted
as a translation operator $\translate{\beta}$ that moves a wave function
a distance $\beta$ without changing its energy. This point of view
will be fruitful in understanding why $b$ and $b^{\dagger}$ fail
to be good operators on the cylinder and torus.

Comparing with the Harmonic Oscillator, the coherent state $\ket{\alpha,\beta}$
now precess in real space whereas the Harmonic Oscillator state precesses
in phase space. We may thus think of the real space probability distribution
$\left|\braket z{\alpha,\beta}\right|^{2}$ in a magnetic field, in
analogy to the phase space quasi-probability distribution of the Harmonic
Oscillator\cite{Schleich_01}, even though the concepts are not mathematically
equivalent. An important difference is that in the Harmonic Oscillator
all states circulate around $\left\langle x\right\rangle =\left\langle p\right\rangle =0$,
whereas in the magnetic field the coherent states may circulate around
any point $\left\langle z\right\rangle =\beta$. This difference introduces
an extra degree of freedom, which will affect the uncertainty relations
\eqref{eq:Uncertianty-relation}. One special uncertainty relation
that will be modified, is between $x$ and $y$, within a given LL.
Because of the vector potential, $y$ will play the role of $p$,
with the existence of the magnetic field. In terms of ladder operators,
the positions operators are 
\begin{eqnarray*}
\hat{x} & = & \frac{\ell}{\sqrt{2}}\left(a+b+a^{\dagger}+b^{\dagger}\right)\\
\hat{y} & = & \frac{\ell}{\i\sqrt{2}}\left(b+a^{\dagger}-a-b^{\dagger}\right).
\end{eqnarray*}
 Within the LLL we define the projected operators as 
\begin{eqnarray*}
\hat{x}_{\mathrm{LLL}} & = & \plll\hat{x}\plll=\frac{\ell}{\sqrt{2}}\left(b+b^{\dagger}\right)\\
\hat{y}_{\mathrm{LLL}} & = & \plll\hat{y}\plll=\frac{\ell}{\i\sqrt{2}}\left(b-b^{\dagger}\right),
\end{eqnarray*}
 and these do not commute, $\left[\hat{x}_{\mathrm{LLL}},\hat{y}_{\mathrm{LLL}}\right]=\i\ell^{2}$.
Thus the product $\sigma_{x}\sigma_{y}$ will be calculated repeatedly
in the coming sections. We will call this measure the \emph{delocalization,}
since $\sigma_{x}\sigma_{y}$ is a measure of the occupied area of
a state. The minimal $\sigma_{x}\sigma_{y}$ delocalization within
a LL will however not be $\frac{\ell^{2}}{2}$ as we would have expected
from the analogy with the Harmonic Oscillator. Instead it will be
$\ell^{2}$, since now there exists two ladder operators that contribute
to both the $x$ and $y$ operators. The easy way to see this is that
$\sigma_{x}^{2}=\left\langle x^{2}\right\rangle -\left\langle x\right\rangle ^{2}$
is different from $\sigma_{x_{\mathrm{LLL}}}^{2}=\left\langle x_{\mathrm{LLL}}^{2}\right\rangle -\left\langle x_{\mathrm{LLL}}\right\rangle ^{2}$,
even within a single LL.

Indeed we see for the coherent states, that $\left\langle x\right\rangle =\Re\left(\beta+\alpha\right)$
and $\left\langle y\right\rangle =\Im\left(\beta-\alpha\right)$ and
$\left\langle x^{2}\right\rangle =\Re^{2}\left(\alpha+\beta\right)+1$
as well as $\left\langle y^{2}\right\rangle =\Im^{2}\left(\beta-\alpha\right)+1$.
Thus when we restore units $\sigma_{x}\sigma_{y}=\ell^{2}$ and these
states saturate the Heisenberg uncertainty relation. The states $\ket{0,\beta}$
are energy eigenstates since $b$ and $b^{\dagger}$ are not present
in the Hamiltonian. These states are thus stationary under time evolution
and represent localized LLL particles. In fact, the state $\ket{0,\beta}$
can be obtained by projecting a spacial delta function $\delta^{\left(2\right)}\left(z-\beta\right)$
onto the LLL such that $\braket z{0,\beta}=\plll\delta^{\left(2\right)}\left(z-\beta\right)$.

On the torus, we would like to perform the same construction as on
the plane, and find localized states within the LLL. Unfortunately
there are no analogues of the $b$ or $b^{\dagger}$ operators here
since we break the rotational invariance. Under this change of geometry,
the $b$ and\textbf{ $b^{\dagger}$} operators are replaced by translation
operators $t_{x}$ and $t_{y}$. These operators have different commutation
relations than $b$,\textbf{ $b^{\dagger}$}. A consequence of this
problem is that $\left\langle z\right\rangle $ is no longer well-defined,
as it will depend on how the torus is parametrized. In fact, already
the cylinder poses a problem, as it has periodic boundary conditions
in one direction. Going to the torus only makes matters worse. In
essence, since rotational invariance is broken down to translational
invariance, another basis needs to be found. On the cylinder the basis
of choice is a linear basis, which respects the geometry of the cylinder.
These states are plane waves in one direction and localized Gaussians
in the other. Unfortunately there is no natural highest weight state,
\emph{i.e. }there is no state $\ket 0$ from which all other states
can be generated, and which is annihilated by the conjugate operator.
We will clarify this as we more thoroughly define the torus. 

If we cannot use the ladder operators, then what strategy can we use?
We choose to project a spacial delta function onto the LLL as a means
to construct coherent states on the toroidal geometry. Our hope is
that $\plll\delta^{\left(2\right)}\left(z-z^{\prime}\right)$ gives
a state that is analogous to $\alpha=0$ and $\beta=z^{\prime}$.
We will also explore an alternative method of explicitly constructing
a family of coherent states.

\section{The Torus Itself\label{sec:The-torus}}

\begin{figure}
\begin{centering}
\begin{tabular}{ccc}
  \includegraphics[width=0.4\columnwidth]{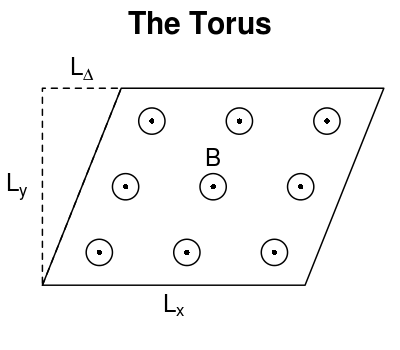} & $\qquad\qquad\qquad$ & 
  \includegraphics[width=0.35\columnwidth]{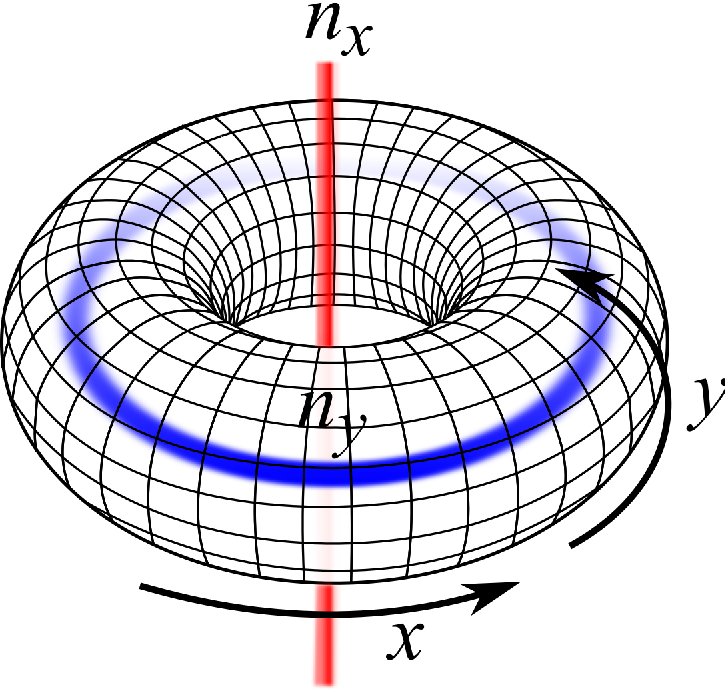}\tabularnewline
$a)$ &  & $b)$\tabularnewline
\end{tabular}
\par\end{centering}

\caption{$a)$The toroidal geometry: Width $L_{x}$, height $L_{y}$, skewness
$L_{\Delta}$. All points on the lattice $\mathbf{r}=n\mathbf{L}_{1}+m\mathbf{L}_{2}=\left(nL_{x}+mL_{\Delta}\right)\hat{\mathbf{x}}+mL_{y}\hat{\mathbf{y}}$
are identified. $b)$ Changing the boundary conditions is equivalent
to inserting fluxes through the two cycles of the torus. As fluxes
$n_{x}$and $n_{y}$ are inserted, the positions of all the states
are transported along the principal directions of the torus. Changing
the boundary conditions by $2\pi$ is equivalent to adding one unit
of flux. \label{fig:The torus}}
\end{figure}

So what do we mean by a torus? In simple words, a torus is a surface
that has periodic boundary conditions in two directions. We can think
of the torus as a doughnut, such as the one depicted in the right
panel of Figure \ref{fig:The torus}, although we should remember
that our torus is locally flat.

Mathematically the torus is characterized by two lattice vectors $\mathbf{L}_{1}=L_{x}\hat{\mathbf{x}}$
and $\mathbf{L}_{2}=L_{\Delta}\hat{\mathbf{x}}+L_{y}\hat{\mathbf{y}}$\nomenclature[3]{$L_y$}{Height  of torus}
\nomenclature[3]{$L_x$}{Width of torus} \nomenclature[3]{$L_\Delta$}{Skewness of torus}
and this geometry is depicted in the left panel of Figure \ref{fig:The torus}.
We should think of $L_{x}$ and $L_{y}$ as the width and height of
the torus respectively whereas $L_{\Delta}$ is the skewed distance
of the torus. Through the surface there is a magnetic field pointing
in the $\hat{\mathbf{z}}$-direction, $\mathbf{B}=B\hat{\mathbf{z}}$.
To describe the magnetic field we will use the Landau gauge $\mathbf{A}=-By\hat{\mathbf{x}}$
such that $\mathbf{B}=\nabla\times\mathbf{A}$. 

The single particle Hamiltonian on the torus is still given by \eqref{eq:General hamiltonian},
and we seek a set of operators that commute with $H$, and can translate
a wave function a distance \textbf{$\mathbf{L}$}. For the free Hamiltonian
$H_{\mathrm{free}}=\frac{\mathbf{p}^{2}}{2m}$, this operator is the
ordinary $t_{\mathrm{free}}\left(\mathbf{L}\right)=e^{\mathbf{L}\cdot\nabla}$,
that has the effect $t_{\mathrm{free}}\left(\mathbf{L}\right)\psi\left(\mathbf{x}\right)=\psi\left(\mathbf{x}+\mathbf{L}\right)$.
In a magnetic field $\left[H,t_{\mathrm{free}}\right]\neq0$,  so
the operator $\translate{\mathbf{L}}$ that translates a wave function
in some direction $\mathbf{L}$ is more complicated than if there
was no magnetic field present. In our specific gauge, the operator
is written as 
\begin{equation}
\translate{\mathbf{L}}=\exp\left[\mathbf{L}\cdot\nabla+\frac{1}{\ell^{2}}\left\{ \mathbf{L}\cdot\i y\hat{\mathbf{x}}-\i\hat{\mathbf{z}}\cdot\left(\mathbf{L}\times\mathbf{r}\right)\right\} \right],\label{eq:translation operator}
\end{equation}
\nomenclature[4]{$\translate{\mathbf L}$}{Translation operator: Sends $\mathbf r\rightarrow \mathbf r+\mathbf L$ and performs gauge transform}where
for clarity the magnetic length $\ell$ has been restored. The first
part of $\translate{\mathbf{L}}$ is the same as for the free Hamiltonian.
The second part of $\translate{\mathbf{L}}$ encodes the gauge transformation
needed to commute with $H$. When convenient, the complex notation
$\translate{\alpha+\i\beta}\equiv\translate{\alpha\hat{\mathbf{x}}+\beta\hat{\mathbf{y}}}$
will be used, and the magnetic length $\ell$ will be set to $\ell=1$.
For translations in the $x$ and $y$ directions, we may evaluate
the effect of the translation operator as $\translate{\alpha\hat{\mathbf{x}}}f\left(x,y\right)=f\left(x+\alpha,y\right)$
and $\translate{\beta\hat{\mathbf{y}}}f\left(x,y\right)=e^{\i\beta x}f\left(x,y+\beta\right)$.
Just as $\hat{x}$ and $\hat{y}$ did not commute on the plane, neither
do translations in different directions. We rather have a magnetic
algebra
\begin{equation}
\translate{\gamma}\translate{\delta}=\translate{\delta}\translate{\gamma}e^{\frac{\i}{2\ell^{2}}\Im\left(\gamma\bar{\delta}\right)},\label{eq:magnetic_algebra}
\end{equation}
such that when translating around a closed loop, we pick up a phase
equal to the area enclosed by the loop. Since the torus has a closed
surface, and there should be no ambiguity in the phase depending on
which side of the loop we choose as the interior, there are constraints
on the area of the torus. Requiring single-valued wave functions in
this way, we find the area of the torus to be $L_{x}L_{y}=2\pi N_{s}$,
where $N_{s}$ is an integer equal to the number of flux quanta that
pierce the torus. We can thus express $L_{x},$ $L_{y}$ and $L_{\Delta}$
in terms of the complex modular parameter $\tau=\frac{1}{L_{x}}\left(L_{\Delta}+\i L_{y}\right)$
and $N_{s}$. 

The periodic boundary conditions are implemented as 

\begin{eqnarray}
\translate{L_{x}}\psi\left(z\right) & = & e^{\i\phi_{1}}\psi\left(z\right)\label{eq:bc_1}\\
\translate{\tau L_{x}}\psi\left(z\right) & = & e^{\i\phi_{2}}\psi\left(z\right),\label{eq:bc_2}
\end{eqnarray}
 where the phase angles $\phi_{i}$ have the physical interpretation
of fluxes threading the two cycles of the torus. The interpretation
is illustrated in Figure \ref{fig:The torus}b. The physical effects
of changing $\phi_{j}$ is that all states on the torus will shift
their positions. By letting $\phi_{j}\rightarrow\phi_{j}+2\pi$, each
state will have transformed into another state a short distance away. 

We now see why the $b$ and $b^{\dagger}$ operators are not useful
on the cylinder and the torus. Imposing periodic boundary conditions
requires that all operators have to commute with $\translate{L_{x}}$
on the cylinder and also $\translate{\tau L_{x}}$ on the torus. Since
the commutator $\left[\translate{L_{x}},b\right]=L_{x}\translate{L_{x}}$
and $\left[\translate{L_{x}},b^{\dagger}\right]=L_{x}\translate{L_{x}}$
are not zero, we find that only the combination $b-b^{\dagger}$ is
allowed on the cylinder. Adding the torus constraint and $\left[\translate{\tau L_{x}},b\right]=\tau L_{x}\translate{\tau L_{x}}$
,$\left[\translate{\tau L_{x}},b^{\dagger}\right]=\bar{\tau}L_{x}\translate{\tau L_{x}}$,
we find that no linear combination of $b$ and $b^{\dagger}$ is allowed
on the torus%
\footnote{In the special case of $\Im\left(\tau\right)=0$ the $b-b^{\dagger}$
operator is still allowed, but then $\tau L_{x}$ and $L_{x}$ are
linearly dependant.%
}. 

As a direct consequence of the imposed boundary conditions on the
torus, not all vectors \textbf{$\mathbf{L}$} are valid in the translation
operator $\translate{\mathbf{L}}$. If we wish to stay within a specific
sector of boundary conditions, then by necessity $\left[\translate{\mathbf{L}},\translate{L_{x}}\right]=\left[\translate{\mathbf{L}},\translate{\tau L_{x}}\right]=0$.
Only a subset of $\translate{\mathbf{L}}$ satisfy this condition.
These translation vectors fall on the lattice $\Gamma=\frac{L_{x}}{N_{s}}n+\frac{L_{x}}{N_{s}}\tau m$
for integers $n$ and $m$. The existence of this sub-lattice necessitates
the introduction of the notation 
\begin{equation}
x_{n}=n\frac{L_{x}}{N_{s}}\qquad\qquad y_{n}=n\frac{L_{y}}{N_{s}}\qquad\qquad\omega_{n}=n\frac{L_{\Delta}}{N_{s}}.\label{eq:x_n,y_n,w_n}
\end{equation}
 Equation \eqref{eq:x_n,y_n,w_n} parametrizes the natural sub-lattice
formed by these translations, that preserve the boundary conditions.
The two operators that map out this lattice are 
\begin{eqnarray}
t_{1}^{n} & \equiv & \translate{x_{n}}\label{eq:t1_def}\\
t_{2}^{m} & \equiv & \translate{\tau x_{m}}=\translate{\omega_{m}+\i y_{m}},\label{eq:t2_def}
\end{eqnarray}
which translate in the two main directions on the sub-lattice.\nomenclature[4]{$t_1$}{Finite translation operator in $x$-direction}\nomenclature[4]{$t_2$}{Finite translation operator in $\tau x$-direction}
In the following we shall fix the boundary conditions to $\phi_{1}=\phi_{2}=0$.
It is at any time possible to restore the generic periodic boundary
conditions of $\phi_{1}$ and $\phi_{2}$ by acting with $\translate{\gamma}$,
where $\gamma=\left(\phi_{1}\tau-\phi_{2}\right)\frac{1}{L_{y}}$.

\section{Basis states}

In the Landau gauge described above, the Hamiltonian for a charged
particle in a magnetic field is expressed as 
\begin{equation}
\hat{H}=\frac{1}{2m}p_{y}^{2}+\frac{1}{2m}\left(p_{x}-eBy\right)^{2}.\label{eq:Landau Hamiltonian}
\end{equation}

On the cylinder, the normalized eigenstates with energy $E_{n}=\hbar\omega\left(n+\frac{1}{2}\right)$
of this Hamiltonian are given by

\begin{equation}
\chi_{n,s}\left(x,y\right)=\frac{1}{\sqrt{L_{x}\sqrt{\pi}}}e^{-\i y_{s}x}H_{n}\left(y-y_{s}\right)e^{-\frac{1}{2}(y-y_{s})^{2}}\label{eq:cylinder_higher_landau_levels}
\end{equation}
 where $H_{n}$ is an Hermite polynomial.\nomenclature[5a]{$\chi_{n,s}$}{Eigenstate of $t_1$ on cylinder}
It is easy to see that $t_{2}^{m}\chi_{n,s}=\translate{y_{m}\hat{\mathbf{y}}}\chi_{n,s}=\chi_{n,s-m}$
so there is no lowest-weight state fulfilling $t_{2}^{-1}\chi_{n,s}=0$.
Since the cylinder has an infinite amount of basis states for both
positive and negative $s$ the bottom will never be reached by the
application of $t_{2}$.

Going from the cylinder to the torus, we must periodize $\chi_{n,s}$
in the $\mathbf{L}_{2}$-direction, as the cylinder functions are
only periodic in the $\mathbf{L}_{1}$-direction. We achieve this
by construction the torus wave function $\eta_{n,s}$, as a linear
combination of the states $\chi_{n,s+kN_{s}}$, $k\in\mathbb{Z}$.
The LLL basis wave functions on the torus are
\begin{equation}
\eta_{s}\left(z\right)=\frac{1}{\sqrt{L_{x}\sqrt{\pi}}}\sum_{t}e^{\i\frac{1}{2}\left(y_{s}+tL_{y}\right)\left(\omega_{s}+tL_{\Delta}\right)}e^{-\i\left(y_{s}+tL_{y}\right)x}e^{-\frac{1}{2}\left(y-y_{s}-tL_{y}\right)^{2}}.\label{eq:eta_expanded_raw}
\end{equation}
 This may be rewritten as 

\begin{eqnarray}
\eta_{s}\left(z\right) & = & \frac{e^{-\frac{1}{2}y^{2}}}{\sqrt{L_{x}\sqrt{\pi}}}\ellipticgeneralized{-\frac{s}{N_{s}}}0{\frac{N_{s}}{L_{x}}z}{N_{s}\tau}.\label{eq:eta_basis_elleipgen}
\end{eqnarray}
\nomenclature[5a]{$\eta_{s}$}{Eigenstate of $t_1$ in LLL on torus}In
equation \eqref{eq:eta_basis_elleipgen}, the generalized quasi-periodic
Jacobi $\vartheta$-function is introduced. (The definition of $\vartheta$
is found in equation \eqref{eq:gen_theta_def} in the Appendix, which
contains a collection of useful formulae related to the Jacobi $\vartheta$-functions.)
From \eqref{eq:eta_basis_elleipgen} is is easy to to see that there
are $N_{s}$ linearly independent basis states, as $\eta_{s+N_{s}}=\eta_{s}$.

The basis $\eta_{s}$ consists of eigenfunctions of $t_{1}$, but
it is also possible to construct eigenfunctions of $t_{2}$ instead.
Since we know that the phase that accompanies commutation of $t_{1}^{n}$
and $t_{2}^{s}$ is $e^{\i x_{n}y_{s}}$, the eigenfunctions of $t_{2}$
can formally be written as $\varphi_{l}\left(z\right)=\frac{1}{\sqrt{N_{s}}}\sum_{s}e^{-\i x_{l}y_{s}}\eta_{s}\left(z\right)$.
Using a transformation property of the $\vartheta$-function under
Fourier sums, \eqref{eq:gen_theta_Fourier_sum}, the eigenfunctions
of $t_{2}$ can immediately be expressed as 
\begin{equation}
\varphi_{l}\left(z\right)=\frac{e^{-\frac{1}{2}y^{2}}}{\sqrt{N_{s}L_{x}\sqrt{\pi}}}\ellipticgeneralized 0{\frac{l}{N_{s}}}{\frac{1}{L_{x}}z}{\frac{\tau}{N_{s}}}.\label{eq:var_psi_basis_final}
\end{equation}
\nomenclature[5a]{$\varphi_{s}$}{Eigenstate of $t_2$ in LLL on torus}A
more physical approach to constructing $\varphi_{l}$ can be taken
by noticing that all the physics should be invariant under the identification
\textbf{$\mathbf{L}_{1}\rightarrow\mathbf{L}_{2}$ }and $\mathbf{L}_{2}\rightarrow-\mathbf{L}_{1}$.
This is equivalent to a rotation of the coordinate system. Seen from
this point of view $\varphi_{l}$ can be obtained from $\eta_{s}$
without the need to explicitly utilize the Fourier summation. This
is done by performing the modular transformation $\tau\rightarrow-\frac{1}{\tau}$,
while letting $z\rightarrow\frac{|\tau|}{\tau}z$, and applying the
appropriate gauge transformation connected with the rotation described
above.

\section{Lattice Coherent States (LCS)\label{sub:LCS}}

An interesting feature of the LLL is that all states in this level
can be written as a Gaussian factor $e^{-\frac{1}{2}y^{2}}$ times
a holomorphic function $\rho\left(z\right)$. Since the torus has
periodic boundary conditions and $\rho\left(z\right)$ is holomorphic,
then $\rho\left(z\right)$ must contain some zeroes, as it would otherwise
be constant. As a consequence of being holomorphic, the function $\rho\left(z\right)$
is also fully determined by the location of these zeroes. We may thus
fully characterize any LLL wave functions by the location of its zeroes.
By choosing these zeroes appropriately, this may allow us, at least
in principle, to engineer states with some desired properties.

In 1985 Haldane and Rezayi proposed a candidate for a localized wave
function. They did so by putting all zeros at the same point\cite{Haldane_85a}.
A wave function with $N_{s}$ fluxes has $N_{s}$ zeroes in the principle
domain, corresponding to the $N_{s}$ linearly independent basis states
at that flux. By fixing the boundary conditions of the wave function,
constraints on the locations of the zeroes are introduced, such that
there are only $N_{s}^{2}$ points where the $N_{s}$-fold zeros can
be. Each of the $N_{s}^{2}$ points corresponds to a wave function.
Since the LLL only can hold $N_{s}$ linearly independent states the
proposed states must be linearly dependent and over-complete. Over-completeness
is nothing troublesome in itself and we have encountered it before,
both in the Harmonic Oscillator and as well in the magnetic field
on the plane. This particular set of states, we shall refer to as,
Lattice Coherent States (LCS)\nomenclature[1]{LCS}{Lattice Coherent States}.
As shall be seen later, it is strictly speaking only in a region around
$\Re\left(\tau\right)=0$ that these states can be considered localized.
As $\tau\rightarrow\tau+\frac{1}{2}$, the LCS goes through a transition
from one localized maxima to two well separated maxima. On a rectangular
($\Re(\tau)=0$) torus, the LCS do approach the expected limit $\sigma_{x}\sigma_{y}=1$
as $N_{s}\rightarrow\infty$. Hence in the thermodynamic limit, the
LCS are likely to be identical to the coherent states on the plane. 

The construction of the LCS rests on the observation that a general
wave function in the LLL on a torus can be written as 
\begin{equation}
\psi\left(z\right)=\mathcal{N}e^{-\frac{y^{2}}{2}}e^{\i kz}\prod_{j=1}^{N_{s}}\elliptic 1{\frac{1}{L_{x}}(z-\xi_{j})}{\tau}\label{eq:haldane_rezaye}
\end{equation}
where $\xi_{j}$ is the position of the $j$:th zero. The function
$\vartheta_{j}$ is defined in equations \eqref{eq:theta_1} to \eqref{eq:theta_4}
in the Appendix, and has the property that $\elliptic 10{\tau}=0$.
By demanding that $\psi\left(z\right)$ obeys periodic boundary conditions
defined by \eqref{eq:bc_1} and \eqref{eq:bc_2}, we get relations
on $k$ and $\bar{\xi}=\frac{1}{N_{s}}\sum_{j}\xi_{j}$. Let us restrict
$\bar{\xi}$ to $\bar{\xi}=x_{1}[m+n\tau]-\frac{L_{x}}{2}[\tau+1]$
and define $z_{j}=\xi_{j}+\frac{1}{2}(1+\tau)L_{x}$. The new variable
$z_{j}$, is the point on the torus where we expect the maximum, of
the coherent state, will be located. This suspicion is based on the
geometric consideration, that if all the zeros $\xi_{j}$ are at the
same point, we will likely find the maximum at the position diametrically
opposed to $\bar{\xi}$. In terms of the new variable $z_{j}$ the
LCS wave function can be brought to the form 
\begin{equation}
\psi_{nm}\left(z\right)=\mathcal{N}_{nm}e^{-\frac{y^{2}}{2}}e^{-\i y_{n}z}\elliptic 3{\frac{\pi}{L_{x}}(z-z_{nm})}{\tau}^{N_{s}},\label{eq:LCS_theta_product}
\end{equation}
where $z_{nm}=x_{m}+x_{n}\tau$.\nomenclature[5]{$\psi_{n,m}$}{LCS wave function}
All the LCS are generated using $t_{1}$ and $t_{2}$ such that $t_{1}^{l}t_{2}^{k}\psi_{n,m}\propto\psi_{n+k,m+l}$.
Using $t_{2}$ and $t_{1}$, the relative normalization of $\psi_{nm}$
and $\psi_{n^{\prime}m^{\prime}}$ can be deduced, by transforming
the different $\psi_{nm}$ into each other. By inspection we see that
$|\mathcal{N}_{nm}|=\mathcal{N}e^{-\frac{y_{n}^{2}}{2}}$, where $\mathcal{N}\equiv\mathcal{N}_{00}$.
We will later, in section \ref{sub: Localization}, calculate $\sigma_{x}\sigma_{y}$
for $\psi_{nm}$ and will then use the expression in \eqref{eq:LCS_theta_product}
as it is well suited for numerical evaluation. However, for analytic
manipulations this is not the most useful way of writing $\psi_{nm}\left(z\right)$.
Furthermore, equation \eqref{eq:LCS_theta_product} also leaves unanswered
the question of how to calculate the normalization $\mathcal{N}_{nm}$. 

To proceed further we need to expand $\psi_{nm}$ in Fourier modes
in such a way that it will resemble \eqref{eq:eta_expanded_raw}.
By hiding parts of the Fourier weight in a constant, $Z_{K}$, we
can write $\psi_{nm}$ as 

\begin{equation}
\psi_{nm}\left(z\right)=\mathcal{N}\sum_{K}Z_{K+n}e^{-\frac{1}{2}\left(y+y_{K}\right)^{2}}e^{\i y_{K}\left(x-x_{m}\right)}e^{\i\frac{1}{2}y_{K}\omega_{K}}\label{eq:LCS_Fourier}
\end{equation}
 where $Z_{K}$ is defined as 
\begin{equation}
Z_{K}=\sum_{\stackrel{\{k_{j}\}=-\infty}{\sum_{j}k_{j}=K}}^{\infty}e^{\i\pi\tau\sum_{j}\tilde{k}_{j}^{2}}.\label{eq:Z_K}
\end{equation}
 The exponential sum runs over $\tilde{k}_{j}$, which is the deviation
from the mean value of $k_{j}$ such that $k_{j}=\frac{K}{N}+\tilde{k}_{j}$.
This constant $Z_{K}$ can, together with a factor $e^{-\i\pi\tau\frac{K^{2}}{N_{s}}}$,
for imaginary $\tau$, be interpreted as the partition function of
$N_{s}$ particles on a circle where the total angular momentum is
constrained to $K$. For our purposes, the most important property
is that $Z_{K+N_{s}}=Z_{K}$. By inspecting \eqref{eq:LCS_theta_product}
and \eqref{eq:LCS_Fourier}, we can fix the relative normalization
as $\mathcal{N}_{nm}=e^{\i y_{n}(x_{m}+\frac{1}{2}x_{n}\tau)}\mathcal{N}$.

In general, when we wish to calculate the overlap between two wave
functions on the torus we might naively think that we would need to
choose a region of integration since the torus only spans a domain
that is $\mathbf{L}_{1}\times\mathbf{L}_{2}$. Because of the periodic
boundary conditions, we are guaranteed that any domain $\mathbf{L}_{1}\times\mathbf{L}_{2}$
will work. Usually what happens is that the $x$-integration gives
a Kronecker $\delta$, that allow us to combine the $y$-integral
from a piecewise incomplete to a complete integral. The LCS are no
exceptions, and with some algebra we find the overlap to be 
\begin{equation}
\braket{\psi_{n^{\prime}m^{\prime}}}{\psi_{nm}}=\sqrt{\mbox{\ensuremath{\pi}}}L_{x}\mathcal{N}^{2}\sum_{l=1}^{N_{s}}Z_{l+n}\bar{Z}_{l+n^{\prime}}e^{\i y_{l}(x_{m^{\prime}}-x_{m})}.\label{eq:overlap_nm_nm_prime}
\end{equation}
 Choosing $m^{\prime}=m$ and $n^{\prime}=n$ we get 
\begin{equation}
\sum_{k=1}^{N_{s}}|Z_{k}|^{2}=\frac{\mathcal{N}^{-2}}{L_{x}\sqrt{\pi}}\label{eq:normalization}
\end{equation}
 that defines the normalization of \eqref{eq:LCS_Fourier}. Although
numerical values for $Z_{k}$ are unknown, we expect that \eqref{eq:overlap_nm_nm_prime}
will resemble a Gaussian function as $N_{s}\rightarrow\infty$. The
argument is most simple for $m\neq m^{\prime}$ and $n=n^{\prime}$.
If the $Z_{k}$ were all constant, then the overlap would reduce to
$\braket{\psi_{nm^{\prime}}}{\psi_{nm}}\propto\delta_{mm^{\prime}}$.
Now, all the terms $Z_{k}$ are not equal, but on the same scale.
This means that $\braket{\psi_{nm^{\prime}}}{\psi_{nm}}$ has a Gaussian
shape centred at $m=m^{\prime}$, that drops to zero as the phases
between the different terms will interfere destructively.

We mentioned earlier that the LCS are over-complete, but that this
does not pose a problem. The reason for this is that we can form a
simple resolution of unity using $\psi_{nm}$, which is 
\begin{equation}
\plll=\frac{1}{N_{s}}\sum_{m,n}\ketbra{\psi_{nm}}{\psi_{nm}}.\label{eq:LCS_r-o-u}
\end{equation}

A detailed proof of \eqref{eq:LCS_r-o-u}, and \eqref{eq:LCS_self_rep_ker},
is found in Ref \cite{Fremling_13a}. When proving \eqref{eq:LCS_r-o-u},
it is essential that $\sum_{m,n}\ketbra{\psi_{nm}}{\psi_{nm}}$ commutes
with both $t_{1}$ and $t_{2}$, since it implies that it is proportional
to $\plll$, the projector onto the LLL .\nomenclature[2]{$\plll$}{Projector onto the LLL}

In Section \ref{sec:CCS} we will see that the Continuous Coherent
States can form a self-reproducing kernel in the LLL. This means that
the coherent states work just like a $\delta$-function, giving $\int d^{2}w\,\varphi_{w}\left(z\right)\psi\left(w\right)=\psi\left(z\right)$,
when $\psi\left(z\right)$ is a LLL wave function. For the LCS, there
exists a similar kernel and it can be formulated as 

\begin{equation}
\phi\left(z\right)=S^{-1}\sum_{m,n=1}^{N_{s}}e^{\i\frac{1}{2}y_{n}\omega_{n}}\psi_{mn}\left(z\right)\phi\left(z_{nm}\right)\label{eq:LCS_self_rep_ker}
\end{equation}
 where $\psi(z)$ again is an arbitrary wave function in the LLL and
$z_{mn}=x_{m}+x_{n}\tau$. Equation \eqref{eq:LCS_self_rep_ker} is
established by first proving the equation on a sub-lattice $z=z_{lp}$,
and then arguing that these are enough points for the formula to be
valid for all $z$ in the fundamental domain. We can view \eqref{eq:LCS_self_rep_ker}
as a map $\plcs$,\nomenclature[2]{$\plcs$}{LCS map to the LLL} from
the space of arbitrary wave functions $\phi(z,z^{\star})$ to the
LLL wave functions. 

In establishing equation \eqref{eq:LCS_self_rep_ker} we have showed
that $\plcs\psi=\plll\psi$, if $\psi$ is in the LLL. We should however
be aware that \eqref{eq:LCS_self_rep_ker} does \emph{not} represent
a true projection operator $\plll$. The reason is that \eqref{eq:LCS_self_rep_ker}
on components that are not in the LLL, does in general not vanish.
This is seen by considering $\delta\left(z-z^{\prime}\right)$, which
has components in all Landau levels, especially in the LLL. It is
obvious that $\plcs\delta\left(z-z^{\prime}\right)$ will be zero,
even though we know that $\delta(z-z^{\prime})$ has components in
the LLL. Thus the effect of $\plcs$ is that the contributions from
non-LLL states precisely cancel the LLL part, except at $z^{\prime}=z_{mn}$
for which the contribution is divergent. This statement can be made
somewhat sharper by considering the simplest case of $N_{s}=1$ --
where each Landau level has only one state, $\eta_{n}$. By a simple
parity argument, we can show that $\plcs\eta_{2n+1}=0$ whereas $\plcs\eta_{2n}\neq0$.
We suspect that the result for $N_{s}=1$ is valid for arbitrary $N_{s},$
meaning that $\plcs\psi=\plll\psi$, if $\psi$ is in an odd numbered
LL (or the LLL), but are otherwise different.

\section{Continuous Coherent states (CCS)\label{sec:CCS}}

In the previous Section we introduced the LCS wave functions as a
candidate for coherent states. One of the problems with the LCS is
that they are only defined on a lattice $z_{nm}$ and not for generic
points on the torus. We would like to have a recipe for constructing
localized wave functions around some other points than the ones allowed
for by the LCS. A natural way of constructing these states would be
to project a $\delta$-function on the LLL. These functions will automatically
fulfil the correct boundary conditions and will hopefully be localized
at the base of the $\delta$-function. We thus define the state 

\begin{equation}
\varphi_{w}\left(z\right)=\plll\delta^{\left(2\right)}\left(z-w\right)\label{eq:definition_CCS}
\end{equation}
 where $w=x^{\prime}+\i y^{\prime}$ as our Continuous Coherent States
(CCS).\nomenclature[1]{CCS}{Continous Coherent States} \nomenclature[5]{$\varphi_w$}{CCS wave function}
The projector $\plll$ can either be expressed in terms of basis states
$\plll=\sum_{s}\ketbra{\eta_{s}}{\eta_{s}}$ or in terms of LCS as
$\plll=\frac{1}{N_{s}}\sum_{mn}\ketbra{\psi_{nm}}{\psi_{nm}}$. Since
$\plll^{2}=\plll$, we directly have that $\braket{\varphi_{w^{\prime}}}{\varphi_{w}}=\varphi_{w}\left(w^{\prime}\right)$
which shows that these states are in general not normalized. From
the definition of $\varphi_{w}\left(z\right)$ also follows a resolution
of unity 
\[
\psi\left(z\right)=\int d^{2}w\varphi_{w}\left(z\right)\psi\left(w\right)
\]
 for states in the LLL and zero otherwise. Whatever form of $\plll$
we choose, we will get the expression 
\begin{eqnarray}
\varphi_{w}\left(z\right) & = & \frac{1}{L_{x}\sqrt{\pi}}\sum_{K,t}e^{-\frac{1}{2}\left(y+y_{K}\right)^{^{2}}}e^{-\frac{1}{2}\left(y^{\prime}+y_{K}+L_{y}t\right)^{^{2}}}\times\nonumber \\
 &  & \qquad\qquad\times e^{-\i y_{K}\left(x^{\prime}-x\right)}e^{-\i\left(\omega_{K}+x^{\prime}\right)L_{y}t}e^{-\i\frac{1}{2}L_{y}L_{\Delta}t^{2}}.\label{eq:CCS-expanded}
\end{eqnarray}
 We need to rewrite this expression in terms of $\vartheta$-functions
as these naturally incorporate the boundary conditions on the torus.
We first identify the sum over $K$ with a $\vartheta$-function.
We get the still quite complicated expression 

\begin{eqnarray*}
\varphi_{w}\left(z\right) & = & \frac{e^{-\frac{1}{2}\left(y^{2}+y^{\prime2}\right)}}{L_{x}\sqrt{\pi}}\sum_{t}e^{\i2\pi t\frac{N_{s}}{2}T^{+}}e^{\i\pi\i\Im\left(\tau\right)\frac{N_{s}}{2}t^{2}}\ellipticgeneralized{-t\frac{N_{s}}{2}}{t\Re\left(\tau\right)}{T^{-}}{\frac{2}{N_{s}}\i\Im\left(\tau\right)}
\end{eqnarray*}
 where $T^{\pm}$ is defined as $T^{\pm}=\frac{1}{L_{x}}\left(z\pm\bar{w}\right)$.

Let us first study the special case of $\Re\left(\tau\right)=0$,
such that we assume that $\tau$ is purely imaginary as $\tau=\i\frac{L_{y}}{L_{x}}$.
Even and an odd number of fluxes $N_{s}$, will have $\varphi_{w}$
with slightly different functional forms. In the even case, $t\frac{N_{s}}{2}$
is an integer and we may ignore it in the argument of $\vartheta$
and directly identify the sum over $t$ as another $\vartheta$-function
such that 

\begin{equation}
\varphi_{w}\left(z\right)=\frac{e^{-\frac{1}{2}\left(y^{2}+y^{\prime2}\right)}}{L_{x}\sqrt{\pi}}\elliptic 3{\frac{N_{s}}{2}T^{+}}{\frac{N_{s}\tau}{2}}\elliptic 3{T^{-}}{\frac{2\tau}{N_{s}}}.\label{eq:CCS_imag_tau_even_Ns}
\end{equation}

\begin{figure}
\begin{centering}
\begin{tabular}{ccc}
\includegraphics[width=0.4\columnwidth]{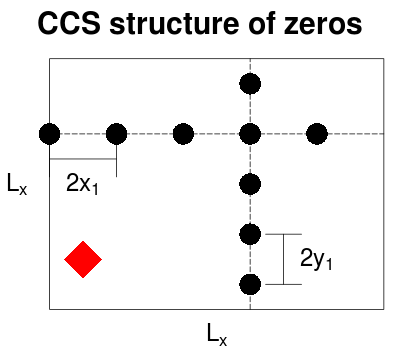} & $\qquad\qquad$ &
\includegraphics[width=0.4\columnwidth]{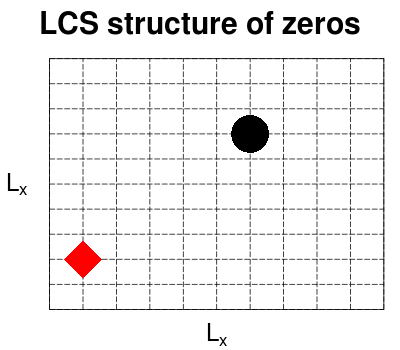}\tabularnewline
$a)$ &  & $b)$\tabularnewline
\end{tabular} 
\par\end{centering}

\caption{Structure of zeros and maxima for $a)$ Continuous Coherent States
$\varphi_{w}$ and $b)$ Lattice Coherent States $\psi_{mn}$. In
both pictures, the red diamond ($\colred{\blacklozenge}$) is centred
at the maximum. The black circles ($\bullet$) represent the locations
of the zeroes. The larger black circle indicates that the zero is
$N_{s}$-fold. On the rectangular ($\Re(\tau)=0$) torus, for $N_{s}$
being an even number, the zeros of the CCS form a ``cross'' centred
over $z=w$. \label{fig:zeros_CCS_LCS} }
\end{figure}

In contrast, if $N_{s}$ is odd, we have to split the sum over $t$
into even and odd terms. We can then identify the even and odd sums
over $t$ separately as $\vartheta$-functions. The resulting wave
function is 

\begin{equation}
\varphi_{w}\left(z\right)=\frac{e^{-\frac{1}{2}\left(y^{2}+y^{\prime2}\right)}}{L_{x}\sqrt{\pi}}\sum_{j=2,3}\elliptic j{T^{+}N_{s}}{2N_{s}\tau}\elliptic j{T^{-}}{\frac{2\tau}{N_{s}}}.\label{eq:CCS_imag_tau_odd_Ns}
\end{equation}

Different functional forms are obtained depending on whether $N_{s}$
is even or odd. This is related to the structure of the zeros. For
an even number of zeros, they will divide into two groups that translate
rigidly under guiding centre translations. For an odd number of zeros,
these rigid translations do not occur. The reason is because there
now exists an extra zero that constrains the movements of the other
zeros. 

For generic values of $\Re\left(\tau\right)$ we can make some progress
by assuming that $\Re\left(\tau\right)$ is a rational number $\Re\left(\tau\right)=\frac{p}{q}$.
For simplicity, we consider only an even number of fluxes. The sum
over $t$ can then be split into smaller pieces $t=k+q\cdot n$ such
that $\sum_{t=-\infty}^{\infty}=\sum_{k=1}^{q}\sum_{q=-\infty}^{\infty}$.
These different sums can separately be identified as $\vartheta$-functions
such that 

\[
\varphi_{w}\left(z\right)=\frac{e^{-\frac{1}{2}\left(y^{2}+y^{\prime2}\right)}}{L_{x}\sqrt{\pi}}\sum_{k=1}^{q}\ellipticgeneralized{\frac{k}{q}}0{\frac{qN_{s}}{2}T^{+}}{\frac{N_{s}}{2}q^{2}\i\Im\left(\tau\right)}\ellipticgeneralized 0{k\frac{p}{q}}{T^{-}}{\frac{2}{N_{s}}\i\Im\left(\tau\right)}.
\]
 For the coherent states with an odd number of fluxes, the situation
is somewhat more complicated, but the logic is the same as for even
fluxes. The precise division of $t$ will now depend on whether $q$
is an even or odd number. 

Since we know where the zeros of the function $\vartheta_{3}$ are
located -- see equation \eqref{eq:ge_theta_zeros} -- we can deduce
the location of the zeros for $\varphi_{w}$ in the case of $\Re\left(\tau\right)=0$
and $N_{s}$ being even. The zeros of $\varphi_{w}$ are at $z=x_{2m+1}-x^{\prime}+\i\left(y^{\prime}+\left(\frac{1}{2}+n\right)L_{y}\right)$
and $z=\i\left(y_{2n+1}-y^{\prime}\right)+x^{\prime}+\left(\frac{1}{2}+m\right)L_{x}$.
The zeros lie on two perpendicular axes intersecting at $z=w+\frac{1}{2}\left(L_{x}+\i L_{y}\right)$,
as depicted in Figure \ref{fig:zeros_CCS_LCS}a. We conclude that
as $\tau$ is transformed away from purely imaginary, the nice linear
pattern formed by the zeros is broken.

\begin{figure}[h]
\begin{centering}
\begin{tabular}{ccc}
\includegraphics[width=0.3\columnwidth]{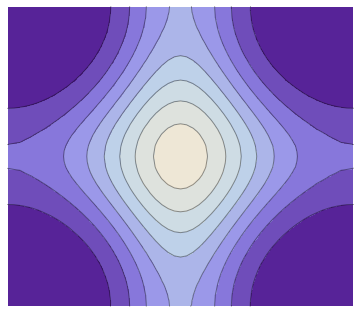} & 
\includegraphics[width=0.3\columnwidth]{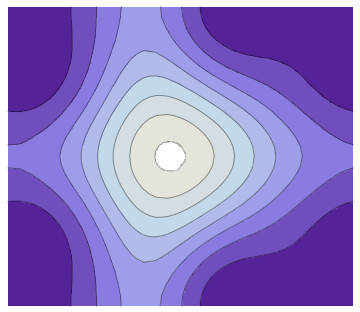} & 
\includegraphics[width=0.3\columnwidth]{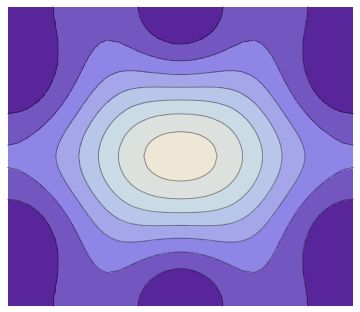}\tabularnewline
$a)$ $w=0$ & $b)$ $w=\frac{1}{4}x_{1}$ & $b)$ $w=\frac{1}{2}x_{1}$\tabularnewline
\includegraphics[width=0.3\columnwidth]{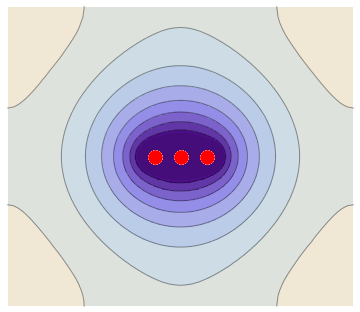} & 
\includegraphics[width=0.3\columnwidth]{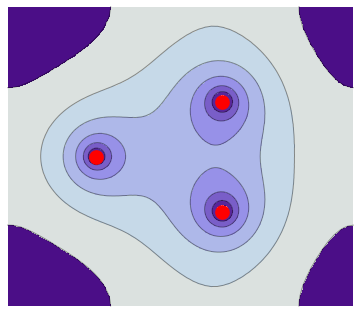} & 
\includegraphics[width=0.3\columnwidth]{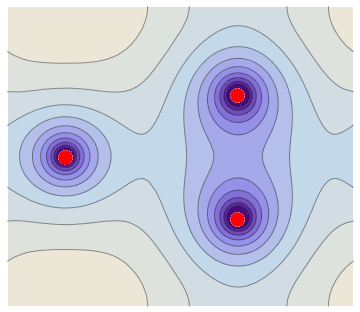}\tabularnewline
\end{tabular}
\par\end{centering}

\caption{\emph{Upper panel} : The spatial profile of CCS at $N_{s}=3$ for
$\tau=\sqrt{\frac{3}{4}}\i$ where $a)$ $w=0$; $b)$ $w=\frac{1}{4}x_{1}$;
and $c)$ $w=\frac{1}{2}x_{1}$. These correspond to a rectangular
lattice where $w$ is moved away from $w=0$. The fundamental domain
is centred around $r=w$. Notice how the spatial profile changes as
$w$ is tuned away from $w=0$. The reason is that the zeros of $\varphi_{w}$
move around.\emph{}\protect \\
\emph{Lower panel}: Positions of the zeros are represented by filled
red circles ($\colred{\bullet}$), for the same values of $w$ as
in the upper panel. Here the domain is fixed with a centre at $r=\frac{1}{2}(1+\tau)x_{1}$
to facilitate the tracking of zeros. \label{fig:spatial_profile_move_r_prime}}
\end{figure}

Further, as we change $w\rightarrow w+\delta w$ half of the zeros
will be propagating in the direction of $\delta w$ while the other
half of the zeros will move in the direction of $-\delta w$. This
behaviour ensures that the boundary conditions are always respected.
The location of the zeroes in $\varphi_{w}$ are $w$ dependent, and
as a consequence the spacial distribution of $\varphi_{w}$ also depends
on $w$. This is illustrated in the upper panel of Figure \ref{fig:spatial_profile_move_r_prime},
for an odd number of particles. Here we plot in the upper panel, the
contours of $|\varphi_{w}|^{2}$ for $N_{s}=3$ and $\tau=\sqrt{\frac{3}{4}}\i$.
The constraint from boundary conditions on the locations of the zeros
is nicely illustrated in the lower panel of Figure \ref{fig:spatial_profile_move_r_prime}.
In the lower panel, we plot $\log\left|\varphi_{w}\right|$ and highlight
the zeros of $\varphi_{w}$ with a filled red circle ($\colred{\bullet}$).
The columns in the Figure are organized such that $a)$ has $w=0$,
$b)$ has $w=\frac{1}{4}x_{1}$ and $c)$ has $w=\frac{1}{2}x_{1}$.
For the upper panel, the fundamental domain is centred at $z=w$ and
in the lower panel the centre of the fundamental domain is at $z=\i\frac{1}{2}\left(1+\tau\right)L_{x}$. 

\begin{figure}
\begin{centering}
\includegraphics[width=0.4\columnwidth]{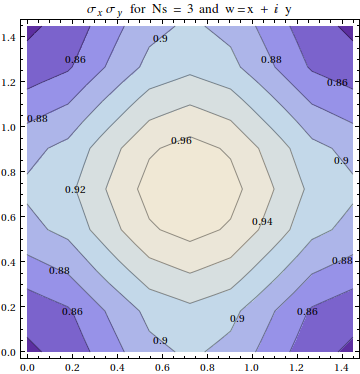} 
\par\end{centering}

\caption{The spatial delocalization of $\varphi_{w}$ is measured as $\sigma_{x}\sigma_{y}$
for $\tau=\i$ with variation of $w$. Darker colour corresponds to
lesser delocalization. The delocalization of $\varphi_{w}$ depends
on $w$. \label{fig:dxdx_of_w}}
\end{figure}

As the probability distribution of $\varphi_{w}$ depends on $w$,
the delocalization $\sigma_{x}\sigma_{y}$ must also depend on $w$.
In Figure \ref{fig:dxdx_of_w}, we can see how $\sigma_{x}\sigma_{y}$
varies as a function of $w$. In the corners, where $w=x_{n}+\tau x_{m}$
the delocalization is at a minimum. In the centre, where $w=x_{n+\frac{1}{2}}+\tau x_{m+\frac{1}{2}}$,
the delocalization is at a maximum.

\section{Localization behaviour of LCS and CCS\label{sub: Localization} }

The previous sections have analysed the LCS and the CCS, that are
candidates for localized wave functions. This chapter will take the
analysis one step further and quantify the spatial delocalization
$\sigma_{x}\sigma_{y}$ for these states. As we have alluded to earlier,
we can not make an analogous calculation to the ones performed in
sections \ref{sub:CS-in-HO} and \ref{sub:CS-in-B-on-plane}, where
we used the ladder operators for an algebraic calculation. It is however
possible to numerically evaluate the $\sigma_{x}\sigma_{y}$ delocalization
using 
\begin{equation}
\left\langle A\left(x,y\right)\right\rangle _{\Omega}=\int\int_{\Omega}dx\, dy\, A\left(x,y\right)\,|f\left(x,y\right)|^{2}\label{eq:expecation-value-integral}
\end{equation}
 where $A\left(x,y\right)$ is some operator and $f\left(x,y\right)$
can be either $\varphi_{w}\left(z\right)$ or $\psi_{nm}\left(z\right)$.
The uncertainty in variable $A$ is defined as $\sigma_{A}^{2}=\left\langle A^{2}\right\rangle -\left\langle A\right\rangle ^{2}$.

Since the mean value is not well defined on a periodic structure,
we need to be careful when we choose the region $\Omega$ in which
we evaluate the integral \eqref{eq:expecation-value-integral}. A
natural choice of the centre $\left(x_{0},y_{0}\right)$ of $\Omega$,
is such that $\left\langle x\right\rangle _{x_{0},y_{0}}=x_{0}$ and
$\left\langle y\right\rangle _{x_{0},y_{0}}=y_{0}$. Here we need
to be careful as there always exists more than one point in any periodic
domain that fulfils $\left\langle x\right\rangle _{x_{0},y_{0}}=x_{0}$
and $\left\langle y\right\rangle _{x_{0},y_{0}}=y_{0}$. To be thorough,
we should choose the point $\left(x_{0},y_{0}\right)$ where $\left\langle x^{2}\right\rangle _{x_{0},y_{0}}$
and $\left\langle y^{2}\right\rangle _{x_{0},y_{0}}$ are minimal. 

\begin{figure}[h]
\begin{centering}
\includegraphics[width=0.5\columnwidth]{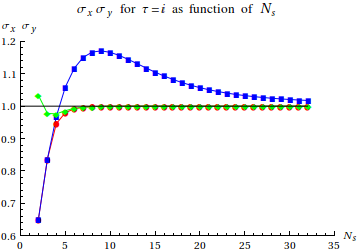} 
\par\end{centering}

\caption{The spatial delocalization of $\varphi_{w}$ and $\psi_{nm}$ measured
as $\sigma_{x}\sigma_{y}$ for $\tau=\i$ and different $N_{s}$.
The colour code is: (\textcolor{red}{Red}) for $\varphi_{w}$ with
$w=x_{m}+\tau x_{n}$; (\textcolor{green}{Green}) for $\varphi_{w}$
with $w=x_{m+\frac{1}{2}}+\tau x_{n+\frac{1}{2}}$; and (\textcolor{blue}{Blue})
for $\psi_{nm}$. The CCS in general displays smaller delocalization
than the LCS but the delocalization of CCS depends on $w$. The LCS
and minimal CCS delocalization are the same for $N_{s}=1,2,3$ for
$\tau=\i$ since their zeros coincide at these fluxes. We have excluded
the point a $N_{s}=1$ since only one state exists at that flux. \label{fig:dxdx_of_Ns}}
\end{figure}

Figure \ref{fig:dxdx_of_Ns} shows how the delocalization depends
on the number of fluxes, $N_{s}$. We will examine the high, $N_{s}\rightarrow\infty$,
and the low, $N_{s}\rightarrow0$, flux limits.

\subsection{The low flux limit $N_{s}=1,\,2,\,3,\,4$.}

In Figure \ref{fig:dxdx_of_Ns}, we first consider the low values
of $N_{s}$, such as $N_{s}=2,\,3,\,4$. We see that $\sigma_{x}\sigma_{y}<1$,
which naively is contradictory to the limit $\sigma_{x}\sigma_{y}\geq1$
set by the Heisenberg uncertainty relation. This issue is resolved
when considering the finite geometry of the torus. This finite geometry
gives an upper bound to how large any delocalization can be. The maximum
delocalization, that of a uniform distribution, has $\sigma_{x}=\frac{1}{\sqrt{3}}\frac{L}{2}$,
where $L$ is the linear width. The main point is that, if the imaginary
part of $\tau$ is far from 1, we can have $L_{x}\ll1$ and $L_{y}\gg1$
such that $\sigma_{y}\approx1$ and $\sigma_{x}\propto L_{x}$. An
illustrative example of this is the basis states $\eta_{s}$, that
have $\sigma_{x}\approx\frac{1}{\sqrt{3}}\frac{L_{x}}{2}$ and $\sigma_{y}\approx\frac{\sqrt{\pi}}{2}$
such that $\sigma_{x}\sigma_{y}\approx\frac{\pi}{2\sqrt{6}}\sqrt{\frac{N_{s}}{\Im\left(\tau\right)}}$,
if $\Im\left(\tau\right)\gtrsim1$. It is obvious that even these
states will, for $\Im\left(\tau\right)$ large enough, violate the
uncertainty relation formulated on the plane. Therefore, it is only
to be expected that the coherent states may violate the uncertainty
relation as well. We can now explain why $\sigma_{x}\sigma_{y}<1$
for the low values of $N_{s}$ in Figure \ref{fig:dxdx_of_Ns}. The
coherent state simply extends over the entire torus such that the
bounds on $\sigma_{x}\sigma_{y}$ do not originate from $\varphi_{w}$,
but rather from the small toroidal size.

\subsection{The Thermodynamic limit $N_{s}\rightarrow\infty$.}

We now inspect Figure \ref{fig:dxdx_of_Ns} again. This time we are
interested in the delocalization, as the number of fluxes increase.
We can see that as $N_{s}\rightarrow\infty$ both the LCS and CCS
approach $\sigma_{x}\sigma_{y}=1$, which is the result on the plane.
It is noteworthy that the CCS converge really fast: At $N_{s}>10$
the CCS have already saturated at the delocalization expected on the
plane, whereas it takes $N_{s}>40$ for the LCS to reach the same
delocalization. For small values of $N_{s}$, the delocalization for
$w\neq0$ can actually be higher than the delocalization of the corresponding
LCS. However already at $N_{s}=9$, the maximum and minimum of the
delocalization are practically indistinguishable for the CCS.

\subsection{Changing the Aspect Ratio of the Torus}

\begin{figure}[h]
\begin{centering}
\begin{tabular}{ccc}
\includegraphics[width=0.4\columnwidth]{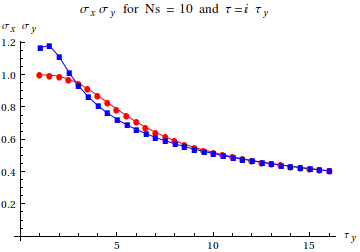} & $\qquad\qquad$ & 
\includegraphics[width=0.4\columnwidth]{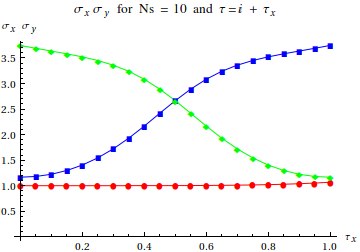}\tabularnewline
$a)$ &  & $b)$\tabularnewline
\end{tabular} 
\par\end{centering}

\caption{The delocalization of CCS (\textcolor{red}{Red}) and LCS (\textcolor{blue}{Blue})
at $N_{s}=10$ as $\tau$ is varied. \emph{$a)$} $\tau=\i\tau_{y}$
where $\tau_{y}$ is varied. The CCS have better delocalization than
the LCS over a region in $\tau$-space around $\tau=\i$\emph{. $b)$}
$\tau=\i+\tau_{x}$, where $\tau_{x}$ is varied. The third line (\textcolor{green}{Green})
shows $\sigma_{x}\sigma_{y}$ calculated for LCS but with $\Omega$
defined by the corners $0$, $L_{x}$,$\tau L_{x}$ and $\tau L_{x}-L_{x}$
instead of $0$, $L_{x}$,$\tau L_{x}$ and $\tau L_{x}+L_{x}$. This
line is included to show that at$\tau\rightarrow\tau+1$ we get localized
states again, but in a different region. \label{fig:dxdy_scan_tau}}
\end{figure}

As we change the aspect ratio of the torus, and let $\Im\left(\tau\right)\rightarrow\infty$,
a similar thing should happen as for low flux. The magnetic length
$\ell=1$, is larger than the linear length of the torus $1>L_{x}$.
As a consequence we expect $\sigma_{x}\sigma_{y}\geq1$ to be violated.
The expected behaviour of $\sigma_{x}\sigma_{y}$ as $\Im\left(\tau\right)\rightarrow\infty$,
is clearly visible in Figure \ref{fig:dxdy_scan_tau}a. There we see
that as $\Im\left(\tau\right)\rightarrow\infty$ then $\sigma_{x}\sigma_{y}\propto\frac{1}{\sqrt{\Im\left(\tau\right)}}$.
We also see that there is a wide region of $\tau$ where the CCS has
lower $\sigma_{x}\sigma_{y}$ than the LCS.

\subsection{Changing the Skewness of the Torus}

\begin{figure}[h]
\begin{centering}
\begin{tabular}{ccccc}
\begin{sideways}
CCS
\end{sideways} & \includegraphics[width=0.2\columnwidth]{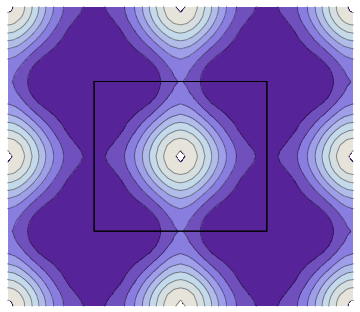}  & 
\includegraphics[width=0.2\columnwidth]{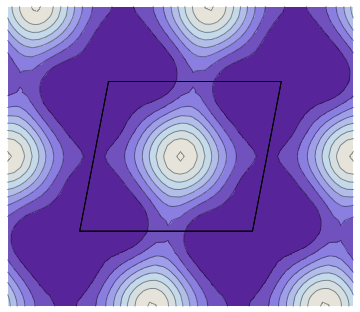}  & 
\includegraphics[width=0.2\columnwidth]{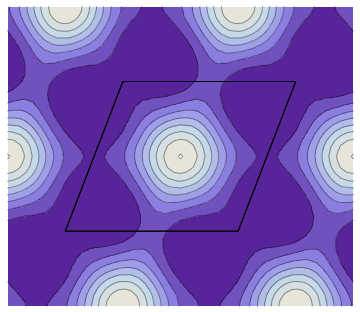}  &
\includegraphics[width=0.2\columnwidth]{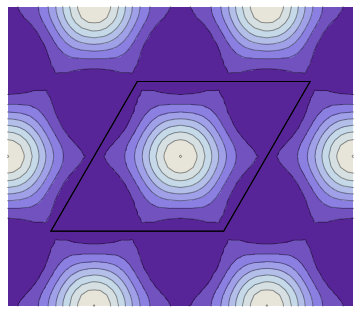}\tabularnewline
 & $a)$ $\tau=\sqrt{\frac{3}{4}}\i$ & $b)$ $\tau=\sqrt{\frac{3}{4}}\i+\frac{1}{6}$ & $c)$ $\tau=\sqrt{\frac{3}{4}}\i+\frac{1}{3}$  & $d)$ $\tau=\sqrt{\frac{3}{4}}\i+\frac{1}{2}$\tabularnewline
\begin{sideways}
LCS
\end{sideways} & \includegraphics[width=0.2\columnwidth]{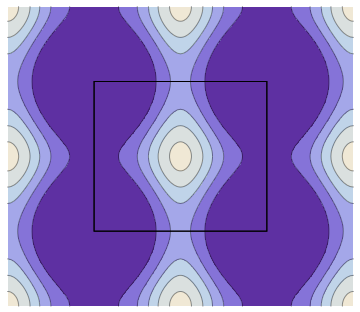} & 
\includegraphics[width=0.2\columnwidth]{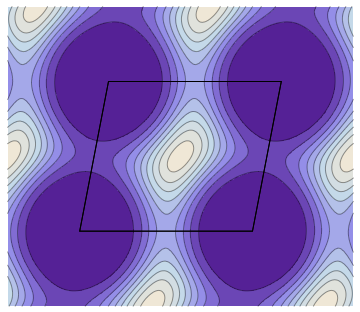} & 
\includegraphics[width=0.2\columnwidth]{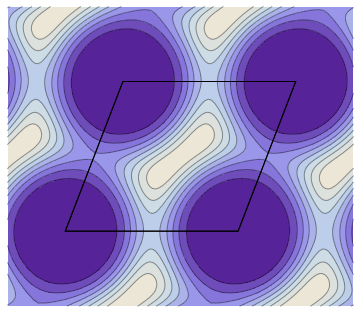}  & 
\includegraphics[width=0.2\columnwidth]{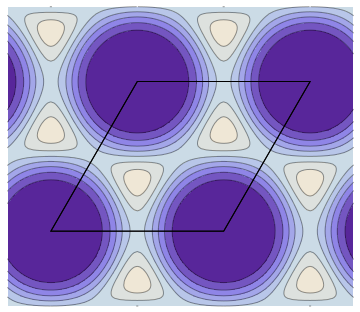} \tabularnewline
\end{tabular} 
\par\end{centering}

\caption{The spatial profile of CCS (\emph{Upper panel}) and LCS (\emph{Lower
panel}) at $N_{s}=4$ for $\tau=\sqrt{\frac{3}{4}}\i+\tau_{x}$ where
$\tau_{x}=0,\frac{1}{6}\frac{1}{3},\frac{1}{2}$. The given $\tau_{x}$
correspond to a rectangular, two general and one triangular lattice.
The black contours show the fundamental domain. The CCS nicely reshape
itself whereas the LCS become massively distorted. Here lighter colour
corresponds to larger values of $|\psi|^{2}$. \label{fig:spatial_profile}}
\end{figure}

Figure \ref{fig:dxdy_scan_tau}b shows what happens if we change the
real part of $\tau$ and keep the imaginary part fixed at $\Im\left(\tau\right)=1$.
It is clear that the CCS wave functions stay localized whereas the
LCS delocalization grows with $\Re\left(\tau\right)$. It is indeed
interesting what goes on here. We can see that since $\sigma_{x}\sigma_{y}$
increases with $\Re\left(\tau\right)$, the state $\psi_{nm}$ is
no longer properly localized with respect to the fundamental region
centred at $z=w$. Comparing $\psi_{nm}$ for $\tau=\i$ and $\tau=\i+1$,
it looks as if the maxima have been shifted by half a period. Instead
of being at $z=z_{nm}$ the maxima of $\psi_{nm}$ is at $z=z_{nm}+\frac{L_{x}}{2}$.
Mathematically this happens since $\elliptic 3{\frac{\pi}{L_{x}}z}{\tau+1}=\elliptic 3{\frac{\pi}{L_{x}}\left(z-\frac{L_{x}}{2}\right)}{\tau}$,
which explains why $\sigma_{x}\sigma_{y}\propto\sqrt{N_{s}}$ at $\tau=1+\i$.
Even more noteworthy, the path the maxima takes as $\tau\rightarrow\tau+1$
is non-trivial. Clearly the maximum is no longer in the centre of
the domain $L_{x}\times\tau L_{x}$ but rather at the centre of the
domain $L_{x}\times\left(\tau+1\right)L_{x}$. Indeed the the maximum
splits up into two separate maxima. This effect is depicted in the
lower panel of Figure \ref{fig:spatial_profile} where we see how
the spatial profile of $\psi_{nm}$ changes as $\tau$ is tuned away
from $\Re(\tau)=0$. As $\tau$ approaches $\Re\left(\tau\right)=1$,
each maximum will recombine with another maximum to finally get the
single maximum localized at $z=z_{nm}+\frac{L_{x}}{2}$ at $\tau=\tau+\i$. 

\begin{figure}[h]
\begin{centering}
\begin{tabular}{ccc}
\includegraphics[bb=0bp 0bp 93bp 115bp]{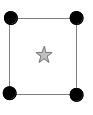}  & $\qquad\qquad$ & 
\includegraphics[bb=0bp 0bp 105bp 115bp]{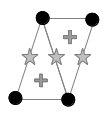}\tabularnewline
$a)$ &  & $b)$\tabularnewline
\end{tabular}
\par\end{centering}

\caption{The analysis for how the zeros and maxima of the of the LCS are located.
$a)$ In the square case $\Re\left(\tau\right)=0$ there is a unique
point farthest from all zeros. This unique point it marked with a
star ({\Large $\star$}). $b)$ For the triangular case $\Re\left(\tau\right)=\frac{1}{2}$
where are two alternatives marked with stars ({\Large $\star$}) and
pluses ({\Large $+$}). We believe that the maxima will always be
at the pluses.\label{fig:LCS_zero_analysis}}
\end{figure}

We believe that we must always have this splitting of the maximum
into two maxima at $\tau=\i\tau_{y}\rightarrow\tau=\i\tau_{y}+\frac{1}{2}$
regardless of the value of $\tau_{y}$. The argument goes as follows
and is illustrated in Figure \ref{fig:LCS_zero_analysis}: For $\tau=\i\tau_{y}$,
the fundamental domain is a square and the point that is the furthest
from all zeros is located at the centre of the domain, assuming the
zeros are at the edges. By symmetry we argue that the maxima is at
this point. At $\tau=\i\tau_{x}+\frac{1}{2}$, the geometry of the
fundamental domain has changed to that of two joined triangles and
there no longer exists a unique point that is farthest from all zeros.
We can therefore not directly determine the position of the maximum.
Instead there exists two possibilities for the maximum, and both of
them will induce multiple maxima. The \emph{first} alternative is
the point at the centre of the fundamental domain, $z=\frac{1}{2}\left(1+\tau\right)L_{x}$,
shown with a star ({\Large $\star$}) in Figure \ref{fig:LCS_zero_analysis}b.
However at this geometry, because of symmetry, an equivalent point
exists also at $z=\frac{1}{2}\tau L_{x}$, also shown with a star.
Because of symmetry, any maximum that is at any of these points must
also be at the other, leading to at least a twofold splitting of the
maxima.

The \emph{second} alternative is to place the maximum somewhere in
one of the triangles, such as in $z_{\mathrm{max}}=\frac{1}{2}L_{x}+\i qL_{y}$
where $0<q<1$ depending on $\tau_{y}$. Due to symmetry there exits
an equivalent point at $z=\left(1+\tau\right)L_{x}-z_{\mathrm{max}}$.
The two points are marked with a plus ({\Large $+$}) in the figure.
Both the alternatives for the location of the maximum results in at
least a twofold maxima. We believe in the latter alternative: Apart
from the suggestive lower panel of Figure \ref{fig:spatial_profile},
the reason is that at the special case of $\tau=\sqrt{\frac{3}{4}}\i+\frac{1}{2}$,
we effectively have a triangular lattice and the maximum at ({\Large $\star$})
would be threefold split.

The conclusion we should draw is that at and around $\tau=\i\tau_{y}+\frac{1}{2}$,
the LCS has a twofold split maximum and can therefore not be localized.
There is reason to believe that the region around $\tau=\i\tau_{y}+\frac{1}{2}$,
that supports double maxima, will be non-vanishing even in the thermodynamic
limit $N_{s}\rightarrow\infty$. This is since the symmetries described
above still exist and the maxima will be sufficiently separated that
small variations in $\tau$ should not affect the stability of the
individual maximum.

It might feel uncomfortable that we violate the Heisenberg uncertainty
relation \eqref{eq:Uncertianty-relation}, when we calculate $\sigma_{x}\sigma_{y}$
on the torus. We should remember, that we violate \eqref{eq:Uncertianty-relation}
because neither $x$ nor $y$ are well defined operators on the torus.
By well defined, we mean that $\braOket{\psi}{\mathcal{O}}{\phi}$
should not depend on the position of the torus fundamental domain
$\Omega$. Given that $\sigma_{x}\sigma_{y}$ is not well defined
on the torus, it is reasonable to question the above analysis all
together. We are however still interested in $\sigma_{x}\sigma_{y}$
for two reasons. First, $\sigma_{x}\sigma_{y}$ is still a measure
of the delocalization of a particle, provided $\Omega$ is chosen
properly. Second, in the thermodynamic limit $N_{s}\rightarrow\infty$,
the area of the torus diverges and we approach the planar limit. In
this limit both $x$ and $y$ recover well defined definitions.


\chapter[Trial Wave Functions from CFT]{Trial Wave Functions from Conformal Field Theory}

So how do the coherent states on the torus relate to conformal field
theory? The connection between the FQHE and CFT lies in the description
of the quasi-particles. It can be shown that the topological information
in a FQH-state can be described using an effective Chern-Simons theory.
The same theory is also conjectured to describe the edge excitations
of the same FQH-state\cite{Wen_92}. It is further conjectured that
trial wave functions with correct topological properties may be extracted
from correlation functions of CFTs with suitable operators\cite{Moore_91}.

The Laughlin wave function at filling fraction $\nu=\frac{1}{q}$
will serve as an introductory example. Simply put, the norm of the
wave function $\left|\psi_{\mathrm{Laughlin}}\right|^{2}$ can be
obtained as a correlation function of a CFT, with primary fields and
a suitable background charge, such that
\[
\left|\psi_{\mathrm{Laughlin}}\right|^{2}\propto\left\langle \mathcal{O}_{\mathrm{bg}}\prod_{i=1}^{N_{e}}V\left(\mathbf{r}_{i}\right)\right\rangle \propto\prod_{i<j}^{N_{e}}\left|z_{i}-z_{j}\right|^{2q}\exp\left\{ -\sum_{i=1}^{N_{e}}\frac{1}{2}\left|z_{i}\right|^{2}\right\} .
\]
 Here $V\left(\mathbf{r}\right)$ is a vertex-operator that represents
an electron at position $\mathbf{r}$. The background operator $\mathcal{O}_{\mathrm{bg}}$
is needed in order to make the whole correlator charge neutral, as
well as to represent the charged atomic background. We now add, that
a many-particle wave function in the LLL can always we written as
$\psi_{\mathrm{LLL}}=\exp\left\{ -\sum_{i=1}^{N_{e}}\frac{1}{4}\left|z_{i}\right|^{2}\right\} \cdot f\left(\left\{ z_{i}\right\} \right)$,
where $f\left(\left\{ z_{i}\right\} \right)$ is a holomorphic function.
This enables us to factorize the correlation function into a holomorphic,
an anti-holomorphic, and a Gaussian part as 
\begin{eqnarray*}
\left\langle \mathcal{O}_{\mathrm{bg}}\prod_{i=1}^{N_{e}}V_{i}\left(\mathbf{r}_{i}\right)\right\rangle  & \propto & \left(\prod_{i<j}^{N_{e}}\left(z_{i}-z_{j}\right)^{q}\cdot\exp\left\{ -\sum_{i=1}^{N_{e}}\frac{1}{4}\left|z_{i}\right|^{2}\right\} \right)\times\\
 &  & \times\left(\prod_{i<j}^{N_{e}}\left(\bar{z}_{i}-\bar{z}_{j}\right)^{q}\cdot\exp\left\{ -\sum_{i=1}^{N_{e}}\frac{1}{4}\left|z_{i}\right|^{2}\right\} \right).
\end{eqnarray*}

Thus, by taking the square root of the correlation function we can
reconstruct the Laughlin wave function \eqref{eq:Lauhlin-wfn}. The
factorization of the correlator if more than symbolic. On the plane
and sphere, the vertex operator $V\left(\mathbf{r}\right)$ can be
split into a holomorphic $V\left(z\right)$ and an anti-holomorphic
part $\bar{V}\left(\bar{z}\right)$, that can be evaluated independently.
On the torus, the separation into holomorphic and anti-holomprphic
is not as clean, as there exists zero-modes, due to the possibility
of winding around the torus handles.

For other states, higher up in the hierarchy the method is the same,
\emph{i.e.} we construct the trial wave function as a correlator of
electron operators. The main difference is that not all electrons
are equivalent. Some electrons will reside in higher LLs, giving rise
to $\bar{z}$ components in the wave function. Under projection the
the LLL these components are transformed into holomorphic derivatives
$\partial_{z}$, acting on the remaining wave function\cite{Girvin_84}. 

In fact, any state within the hierarchy that can be formed through
condensation of quasi-particles, can be expressed in a similar way
as $\psi_{\mathrm{Laughlin}}$, but with some added complexity. First,
there is usually more than one type of electron operator $V^{\left(\alpha\right)}$.
Second, there are external derivatives $\partial_{z}$ acting on the
correlator. Third, the whole wave functions needs to be antisymetrized
explicitly, since all electrons are not treated on equal footing.
Taking all of the above considerations into account, the square of
the Hierarchy wave function may be written symbolically as

\begin{equation}
\left|\psi_{\mathrm{Hierarchy}}\right|^{2}\propto\mathcal{A}\left\{ \left[\mbox{Derivatives}\right]\left\langle \mathcal{O}_{\mathrm{bg}}\prod_{\alpha}\prod_{i_{\alpha}\in I_{\alpha}}\hat{V}^{\left(\alpha\right)}\left(\mathbf{r}_{i_{\alpha}}\right)\right\rangle \right\} .\label{eq:Hierarchy_correlator}
\end{equation}
 In the above equation, $\mathcal{A}$ denotes antisymetrization over
electrons, and $\prod_{\alpha}$ partitions the electrons into different
inequivalent sets. The derivatives come about because some of the
electron operators $V^{\left(\alpha\right)}\left(z\right)$ are describing
the hierarchical fusing of quasi-particles and electrons. This manifests
itself though the appearance of derivatives in the vertex operators
$V^{\left(\alpha\right)}\left(z\right)=\partial_{z}^{\alpha-1}\hat{V}^{\left(\alpha\right)}\left(z\right)$.

On the torus, we usually use a different gauge, such that 
\[
\psi_{\mathrm{LLL}}=\exp\left\{ -\sum_{i=1}^{N_{e}}\frac{1}{2}y_{i}^{2}\right\} \cdot f\left(\left\{ z_{i}\right\} \right).
\]
Also, we expect a ground state degeneracy that can be divided by the
denominator $q$, of the filling fraction $\nu=\frac{p}{q}$\cite{Haldane_85b}.
For the simplest abelian FQH states the degeneracy is exactly $q$.
The ground state degeneracy makes the analysis of the CFT construction
more involved but the basic idea is the same. We are still using electron
operators $V_{i}^{\left(\alpha\right)}\left(\mathbf{r}_{i}\right)$,
such that the correlator is calculated using \eqref{eq:Hierarchy_correlator}.
The problem arises here after we have constructed the correlator for
the many-particle wave function, and need to act with the external
derivatives. These derivatives do not respect the boundary conditions
for the single particle states, and we are interested in how to modify
these on the torus. We thus seek operators that preserve the boundary
conditions, and give us a $q$-fold set of trial wave functions. A
natural candidate is $\plll\partial_{x}$, the derivative projected
to the LLL. In the coming chapter this projection will give some insight
to what operator should be used on the torus. We find that after projection
the derivative turns into a linear combination of translation operators
as 
\[
\plll\partial_{z}=\sum_{l}a_{l}t_{1}^{l}\equiv\mathcal{D}.
\]
 For a many-particle state a product of derivatives would be 
\[
\plll\prod_{i}\partial_{z_{i}}=\prod_{i}\mathcal{D}_{i}.
\]
 We can however show that $\prod_{i}\mathcal{D}_{i}$ does \emph{not}
commute with $T_{2}^{k}=\prod_{j}t_{2,j}^{k}$ for any value of the
power $k$.\nomenclature[4]{$T_1$}{Many body finite translation operator in $x$-direction: $T_1=\prod_jt_{1,j}$}\nomenclature[4]{$T_2$}{Many body finite translation operator in $\tau x$-direction: $T_2=\prod_jt_{2,j}$}
This means that $\prod_{i}\mathcal{D}_{i}$ changes the quantum numbers
in such a way that the $q$-fold degeneracy is lost. The only terms
in $\prod_{i}\mathcal{D}_{i}$ that do not change the $q$-fold degenerate
subspace of the wave functions, are are on the form $T_{1}^{k}=\prod_{i}t_{1,i}^{k}$.
These are the terms that where used in Ref. \cite{Hermanns_08} when
they first addressed this problem. 

A related problem is connected with the description of hole-condensates,
briefly mentioned in Section \ref{sub:Laughlin-Hierarchy}, where
terms that include powers of $\bar{z}$ are generated. These anti-holomorphic
terms cause the wave function to be located in higher Landau levels,
such that it has to be projected down to the lowest one. In symmetric
gauge, this is readily done by the substitution $\bar{z}\rightarrow\partial_{z}$.
On the torus, in Landau gauge, the prescription $\bar{z}\rightarrow\partial_{z}$
will not work, and it is not clear what should replace it. The idea
in this thesis is to use coherent states as a way to project the wave
functions to the LLL, by interpreting the correlators as coefficients
for the coherent state wave functions\cite{Soursa_11a,Soursa_11b}.
We will in the following sections treat the anti-holomorphic $\bar{z}$
components, and the external derivatives $\partial_{z}$, in order.

\section[A Concrete Example: The Modified Laughlin State]{A Concrete Example: \protect \\
The Modified Laughlin State}

Let us take a concrete example of how the coherent state kernel can
be used to project onto the LLL. We here consider states that contains
both $z$ and $\bar{z}$ components, but no derivatives. An example
of such a state, is a modification of the Laughlin state at $\nu=\frac{1}{q}$,
first proposed by Girvin and Jach\cite{Girvin_84} on the plane. They
proposed a short distance modification 
\[
e^{-\frac{q+2p}{4q}\sum_{j}\left|z_{j}\right|^{2}}\prod_{i<j}\left(\bar{z}_{i}-\bar{z}_{j}\right)^{p}\left(z_{i}-z_{j}\right)^{q+p},
\]
 to the Laughlin wave function \eqref{eq:Lauhlin-wfn}. After convolution
with the coherent state kernel, the LLL projection was cast in the
form 
\begin{equation}
e^{-\frac{1}{4}\sum_{j}\left|z_{j}\right|^{2}}\prod_{i<j}\left(\partial_{z_{i}}-\partial_{z_{j}}\right)^{p}\left(z_{i}-z_{j}\right)^{q+p},\label{eq:Girvin-Jash-plane-projected}
\end{equation}
 where the $\bar{z}_{i}$ terms were replaced by derivatives $\partial_{z_{i}}$\cite{Girvin_84}.
This state can be obtained using CFT correlators, either on the plane
or the torus, as a representation in the space of coherent states.
The wave function obtained from the CFT correlator $\psi\left(z,\bar{z}\right)$,
will in general contain both holomorphic coordinates $z$, as well
anti-holomorphic coordinates $\bar{z}$. This wave functions should
not be interpreted directly in the the electron coordinate basis $\left(z,\bar{z}\right)$,
but rather in the over complete basis of coherent states $\varphi_{\xi}$.
Interpreting $\psi\left(z,\bar{z}\right)$ as the weight for the CS
state $\varphi_{\xi}\left(z\right)$, is the same as projecting onto
the LLL, such that 
\begin{equation}
\plll\psi\left(z,\bar{z}\right)=\int d^{2}\xi\,\,\,\psi\left(\xi\right)\cdot\varphi_{\xi}\left(z\right).\label{eq:PLLL_equiv_CS_convolution}
\end{equation}
As a direct consequence of \eqref{eq:PLLL_equiv_CS_convolution},
it can be proven that the boundary conditions that hold for $\psi\left(\xi\right)$,
will also hold for $\plll\psi\left(z,\bar{z}\right)$. The relation
before and after projection is trivial, because the magnetic translation
operators commutes with $\plll$. This fact is easily realized since
$\plll=\prod_{n=1}\left(1-\frac{a^{\dagger}a}{n}\right)$ contains
only the operator $a$, and $\translate{\alpha}=e^{\alpha b^{\dagger}-\bar{\alpha}b}$
contains only the operator $b$. These two operators, $\plll$ and
$\translate{\alpha}$, therefore commute, by the definition of $a$
and $b$. Thus, assuming periodic boundary conditions, such as $\translate{\mathbf{L}_{k}}\psi\left(\xi\right)=e^{\i\phi_{k}}\psi\left(\xi\right)$,
then the same conditions apply for $\translate{\mathbf{L}_{k}}\plll\psi\left(z,\bar{z}\right)=e^{\i\phi_{k}}\plll\psi\left(z,\bar{z}\right)$.

We will now use the basis of coherent states, to find a torus torus
version of \eqref{eq:Girvin-Jash-plane-projected}, with well defined
boundary conditions. For the full procedure of constructing wave functions
on the torus, see Ref. \cite{Hermanns_08}. What is important is that
we have the same kind of short distance behaviour on the torus, as
on the plane. We must also obtain the same $q$-fold degeneracy as
for the ordinary torus Laughlin state. All states with the short distance
behaviour $z_{ij}^{q+p}\bar{z}_{ij}^{p}$ can be calculated in the
same formalism. Hence we will deal with them simultaneously. On the
torus, the polynomial Jastrow factor $\prod_{i<j}\left(z_{i}-z_{j}\right)^{q+p}\left(\bar{z}_{i}-\bar{z}_{j}\right)^{p}$
must be replaced by Jacobi $\vartheta$-functions as 
\[
\prod_{i<j}\elliptic 1{\frac{1}{L_{x}}\left(z_{i}-z_{j}\right)}{\tau}^{p+q}\elliptic 1{-\frac{1}{L_{x}}\left(\bar{z}_{i}-\bar{z}_{j}\right)}{-\bar{\tau}}^{p}.
\]
 The Gaussian factor must be present, and there will also be a centre
of mass component, that is absent on the plane. All these pieces fall
into place as we construct the correlator $\left\langle \prod_{i=1}^{N_{e}}V\left(z_{i},\bar{z}_{i}\right)\mathcal{O}_{\mathrm{bg}}\right\rangle $,
where \textbf{$\mathcal{O}_{\mathrm{bg}}$ }is a suitably chosen background
charge and 
\begin{equation}
V\left(z,\bar{z}\right)=e^{\i\sqrt{p+q}\varphi_{1}\left(z,\bar{z}\right)+\i\sqrt{p}\varphi_{2}\left(z,\bar{z}\right)}\label{eq:CFT_vertex}
\end{equation}
 represents an electron. On the torus, the electron operator $V\left(z,\bar{z}\right)$
\emph{can not} be spit into a holomorphic part $V\left(z\right)$,
and an anti-holomorpic part $\bar{V}\left(\bar{z}\right)$, therefore
the correlator of the full vertex operator needs to be computed. The
two fields $\varphi_{1}$ and $\varphi_{2}$ are two decoupled compactified
boson fields with radius $R_{1}$ and $R_{2}$, respectively. The
correlator is computed as a sum over conformal blocks, $\left\langle \prod_{i=1}^{N_{e}}V\left(z_{i},\bar{z}_{i}\right)\mathcal{O}_{\mathrm{bg}}\right\rangle =N\left(\tau\right)\sum_{E_{1},E_{2}}\psi_{E_{1},E_{2}}\bar{\psi}_{\bar{E}_{1},\bar{E}_{2}}$.
The sum over $E_{1}$ and $E_{2}$ runs over the points $E_{j}=\frac{e_{j}}{R_{j}}+\frac{m_{j}R_{j}}{2}$
and $\bar{E}_{j}=\frac{e_{j}}{R_{j}}-\frac{m_{j}R_{j}}{2}$ and $e_{j},m_{j}\in\mathbb{Z}$.
The chiral (and anti-chiral) terms in the sum can be written as 
\[
\psi_{E_{1},E_{2}}=e^{-\frac{1}{2}\frac{q+p}{q}\sum_{i}y_{i}^{2}}\prod_{i<j}\elliptic 1{z_{ij}}{\tau}^{q+p}\elliptic 1{-\bar{z}_{ij}}{-\bar{\tau}}^{p}\mathcal{F}_{E_{1},E_{2}}\left(Z|\tau\right).
\]
 where $z_{ij}=\frac{1}{L_{x}}\left(z_{i}-z_{j}\right)$ and $Z=\frac{1}{L_{x}}\sum_{j}z_{j}$
is the centre of mass variable. The centre of mass function $\mathcal{F}_{E_{1},E_{2}}$
is given by 
\[
\mathcal{F}_{E_{1},E_{2}}\left(Z|\tau\right)=e^{\i\pi\left[\tau E_{1}^{2}-\bar{\tau}E_{2}^{2}\right]}e^{2\pi\i\left[E_{1}\sqrt{(}q+p)Z-E_{2}\sqrt{p}\bar{Z}\right]}.
\]

The QH wave functions are constructed using only the chiral parts
$\psi_{E_{1},E_{2}}$ of the correlator. Therefore, we now seek linear
combinations of $\psi_{E_{1},E_{2}}$ that have good single- and many-body
electron properties. The singe-body properties sought are formulated
in terms of well-defined periodic boundary conditions under $t_{1}^{N_{s}}$
and $t_{2}^{N_{s}}$. Applying these operators on $\psi_{E_{1},E_{2}}$
yields 
\[
t_{1}^{N_{s}}\psi_{E_{1},E_{2}}=\left(-1\right)^{N_{e}-1}e^{2\pi\i\left[E_{1}\sqrt{q+p}-E_{2}\sqrt{p}\right]}\psi_{E_{1},E_{2}}
\]
 and 

\[
t_{2}^{N_{s}}\psi_{E_{1},E_{2}}=\left(-1\right)^{N_{e}-1}\psi_{E_{1}+\sqrt{q+p},E_{2}+\sqrt{p}}.
\]
 In order to obtain well-defined phases under $t_{1}^{N}$ and $t_{2}^{N}$,
we must choose a linear combination on the form 
\begin{equation}
\phi_{\Gamma,t}=\sum_{k}e^{\i tk}\psi_{\Gamma+k\left(\sqrt{q+p},\sqrt{p}\right)}.\label{eq:Mod-Laugh-generic-sum}
\end{equation}
Here $\Gamma=\left(\Gamma_{1},\Gamma_{2}\right)$ is the offset from
the origin, of the space spanned by $E_{1}$ and $E_{2}$. Depending
on the choice of $\Gamma$, the state $\phi_{\Gamma,t}$ will have
different single-body boundary conditions. Specifying $\Gamma$ also
selects the many-body momentum state of $\phi_{\Gamma,t}$. In what
follows we will investigate the consequence of choosing different
$\Gamma$. We will find that different choices of $\Gamma$ only amounts
to choosing different boundary conditions, and changing the coordinate
system. In order to see this we will make a rather unusual division
of the Gaussian factor as $e^{-\frac{q+2p}{2q}\sum y_{j}^{2}}=e^{-\frac{q+2p}{2q}\sum\tilde{y}_{j}^{2}}e^{-\pi\left(q+2p\right)\frac{L_{x}}{L_{y}}Y^{2}}$
. Here we have rewritten $y_{j}=\tilde{y}_{j}+\frac{YL_{x}}{N_{e}}$
in terms of the centre of mass coordinate $Y=\Im\left(Z\right)$,
and the relative coordinate $\tilde{y}_{j}$. The relative coordinate
is of curse chosen such that $\sum_{j}\tilde{y}_{j}=0$. By performing
the sum over $k$ in \eqref{eq:Mod-Laugh-generic-sum}, we construct
the full centre of mass function:
\begin{eqnarray*}
\mathcal{H}_{\Gamma,t} & = & e^{-\pi\left(q+2p\right)\frac{L_{x}}{L_{y}}Y^{2}}\sum_{k}e^{\i tk}\mathcal{F}_{\Gamma+k\left(\sqrt{q+p},\sqrt{p}\right)}\\
 & = & e^{-\pi\left(q+2p\right)\frac{L_{x}}{L_{y}}Y^{2}}e^{2\pi\i\left(\Gamma_{1}\sqrt{q+p}Z-\Gamma_{2}\sqrt{p}\bar{Z}\right)}e^{\i\pi\left(\tau\Gamma_{1}^{2}-\bar{\tau}\Gamma_{2}^{2}\right)}\times\\
 &  & \quad\times\ellipticgeneralized 0t{\left(q+p\right)Z-p\bar{Z}+\tau\Gamma_{1}\sqrt{q+p}-\bar{\tau}\Gamma_{2}\sqrt{p}}{\tau\left(q+p\right)-\bar{\tau}\left(q\right)}
\end{eqnarray*}
 Note that $\mathcal{H}_{\Gamma,t}$ is the complete centre of mass
function. The expression for $\mathcal{H}_{\Gamma,t}$ looks rather
nasty but can be reformulated. First parametrize $\Gamma$ as $\Gamma=r\kappa+s\lambda$,
where $\kappa=\left(\sqrt{q+p},\sqrt{p}\right)$, and $\lambda=\left(\sqrt{p},-\sqrt{q+p}\right)$,
are two orthogonal vectors. Then $\mathcal{H}_{\Gamma,k}$ simplifies
to

\begin{eqnarray}
 &  & \mathcal{H}_{t,r,s}\left(Z,\tau\right)=e^{-\i2\pi rt}e^{2\pi\i sqX}e^{-\pi\left(q+2p\right)\frac{L_{x}}{L_{y}}Y^{2}}\times\nonumber \\
 &  & \qquad\times\ellipticgeneralized r{t+sq\Re\left(\tau\right)}{\left(q+p\right)Z-p\bar{Z}}{\tau\left(q+p\right)-\bar{\tau}p}.\label{eq:Full-center-of-mass-factor}
\end{eqnarray}
 To reach \eqref{eq:Full-center-of-mass-factor}, we also rescaled
$s\rightarrow\frac{qs}{2\sqrt{p}\sqrt{p+q}}$, and redefined 
\[
\mathcal{H}_{t,r,s}\rightarrow e^{-\i\pi s^{2}q^{2}\left[\frac{\tau}{p+q}-\frac{\bar{\tau}}{p}\right]}\mathcal{H}_{t,r,s}.
\]
Although it might not be apparent, the parameter $s$ is related to
a change in coordinates, and therefore to a gauge transformation.
This can be seen by shifting $Z\rightarrow Z+\i s\Im\left(\tau\right)$.
Under this change, the $sq\Re\left(\tau\right)$ term is shifted away,
yielding 

\begin{eqnarray}
\mathcal{H}_{t,r,s}\left(Z+\i s\Im\left(\tau\right),\tau\right) & = & e^{-\i\pi s^{2}\Re\left(\tau\right)p}\mathcal{H}_{t,r+s,0}\left(Z,\tau\right).\label{eq:COM_term}
\end{eqnarray}
 This expression demonstrates that $s$ is related to a change of
the origin of the coordinate system, by $y_{j}\rightarrow y_{j}+\frac{sL_{y}}{N_{e}}$. 

Equation \eqref{eq:COM_term} demonstrates that the different choices
of $\Gamma$ are all related. We may therefore choose $\Gamma$ to
our convenience. The simplest choice $\Gamma=r\left(\sqrt{q+p},\sqrt{p}\right)$,
amounts to $s=0$ in \eqref{eq:Full-center-of-mass-factor}. Under
the choice $s=0$, the full modified many-body Laughlin wave function
is\nomenclature[5]{$\psi_n^{(q,p)}$}{Modified Laughlin wave funciton av $\nu=\frac 1 q$}
\begin{eqnarray}
\psi_{t,r}^{\left(q,p\right)} & = & e^{-\frac{1}{2}\frac{q+p}{q}\sum_{i}y_{i}^{2}}\prod_{i<j}\elliptic 1{z_{ij}}{\tau}^{q+p}\elliptic 1{-\bar{z}_{ij}}{-\bar{\tau}}^{p}\times\nonumber \\
 &  & \times\ellipticgeneralized rt{\left(q+p\right)Z-p\bar{Z}}{\tau\left(q+p\right)-\bar{\tau}p}.\label{eq:Full-Modified-Laughlin}
\end{eqnarray}
The boundary conditions under single-particle translations can readily
be found to be

\begin{eqnarray*}
t_{1}^{N_{s}}\psi_{t,r} & = & \left(-1\right)^{N_{e}-1}e^{2\pi\i rq}\psi_{t,r}\\
t_{2}^{N_{s}}\psi_{t,r} & = & \left(-1\right)^{N_{e}-1}e^{-2\pi\i t}\psi_{t,r}.
\end{eqnarray*}
 By fixing $r$ and $t$ we can specify the single-particle boundary
conditions. We see that there is a $q$-fold freedom in choosing the
value of $r$. This freedom is related to the $q$ different possible
many-body states. Under many-body translations $T_{1}=\prod_{j}t_{1,j}$
and $T_{2}=\prod_{j}t_{2,j}$ the state $\psi_{r,t}$ transforms as
\begin{eqnarray*}
T_{1}\psi_{t,r} & = & e^{2\pi\i r}\psi_{t,r}\\
T_{2}\psi_{t,r} & = & e^{-\i2\pi\frac{t}{q}}\psi_{t,r+\frac{1}{q}}.
\end{eqnarray*}
 These two expressions confirm that $T_{2}^{q}\psi_{t,r}=e^{-\i2\pi t}\psi_{t,r}$,
such that a maximal set of mutually commuting many-body operators
constitute: $H$, $T_{1}$ and $T_{2}^{q}$. Here $H$ is the full
interacting many-body Hamiltonian.

By arriving at \eqref{eq:Full-Modified-Laughlin}, we have succeeded
in formulating a torus version of \eqref{eq:Girvin-Jash-plane-projected}
with well-defined boundary conditions. However, we can not at this
time analytically project $\psi_{t,r}$ to the LLL, as an analogue
of the trick $\bar{z}\rightarrow\partial_{z}$ lacking, and the $\vartheta$-factors
make analytical attempts difficult. In Section \ref{sub:Num-eval-mod-Lauglin}
we will numerically evaluate \eqref{eq:Full-Modified-Laughlin}, but
we will first perform a small sanity check.

To check that the LLL component of $\psi^{\left(q,p\right)}$ depends
on $p$, we analyse the special case of just one particle ($N_{e}=1$)
and $\phi_{1}=\phi_{2}=0$. With only a single particle, there is
no Jastrow factor and the modified Laughlin wave function is given
by
\begin{equation}
\psi_{n}^{\left(q,p\right)}\left(Z,\tau\right)=\mathcal{N}e^{-\frac{q+2p}{2\pi q}y^{2}}\ellipticgeneralized{-\frac{n}{q}}0{\left(q+p\right)Z-p\bar{Z}}{\tau\left(q+p\right)-\bar{\tau}p},\label{eq:Single-particel-laughlin-mod}
\end{equation}
 where $\mathcal{N}$ is a normalization. It is straight forward to
show that the proper normalization of $\psi_{n}$ is $\mathcal{N}^{2}=L_{x}\sqrt{\frac{q\pi}{q+2p}}$.
From there, the overlap with the LLL basis states are calculated to
be 
\begin{equation}
\braket{\eta_{n}}{\psi_{m}}=\delta_{nm}\left(\frac{1+2x}{1+2x+x^{2}}\right)^{\frac{1}{4}},\label{eq:overlap-laughlin-single-particle}
\end{equation}
where $x=\frac{p}{q}$. Equation \eqref{eq:overlap-laughlin-single-particle}
shows that $\braket{\eta_{n}}{\psi_{n}}=1$ when $p=0$, and that
$\braket{\eta_{n}}{\psi_{n}}\rightarrow0$ as $p\rightarrow\infty$.
We thus conclude that larger deformations of the Laughlin wave function
(larger $p$), has smaller weight in the LLL.

\section{Numerical evaluation of $\psi^{\left(q,p\right)}$\label{sub:Num-eval-mod-Lauglin}}

Remember that just because the wave functions $\psi_{n}^{\left(q,p\right)}$
are not entirely in the LLL, this does not mean that they are ill-suited
trial wave functions. In their original work, Girvin and Jach noted
that the Laughlin state $\psi^{(q,0)}$ could be improved by considering
components with $p\neq0$. On the torus, the same thing is observed.
We have numerically compared $\psi^{\left(q,p\right)}$ and the ground
state for the Coulomb interaction, by projection $\psi^{\left(q,p\right)}$
on the many-body basis states, in the LLL. Doing so we find, as we
expect, that $\psi^{(q,p)}$ is not entirely in the LLL, for $p\neq0$,
but that the projected wave functions $\plll\psi^{(q,p)}$ still has
good overlap with the Coulomb ground state.

The projection of $\psi^{\left(q,p\right)}$ on the LLL is performed
using Monte Carlo with importance sampling, and the procedure works
as follows: Using $\psi^{\left(q,p\right)}$ as the generating function,
$N$ sets of electron coordinates are chosen using the Metropolis-Hastings
algorithm\cite{Metropolis_53}. Then, $\psi^{\left(q,p\right)}$ as
well as the many-body momentum basis states are evaluated to get $N$
data points. The overlap with $\psi^{\left(q,p\right)}$, and each
basis state $\phi_{s}$ is computed as

\[
\braket{\psi^{\left(q,p\right)}}{\phi_{s}}=\frac{1}{\sqrt{\mu\nu}}\frac{1}{Z_{N}}\sum_{i=1}^{N}\frac{\bar{\psi}^{\left(q,p\right)}\left(x_{i}\right)\phi_{s}\left(x_{i}\right)}{p\left(x_{i}\right)},
\]
 where $p\left(x\right)=\left|\psi^{\left(q,p\right)}\left(x\right)\right|^{2}$
is the probability distribution. The normalizing terms are given by 

\[
Z_{N}=\sum_{i=1}^{N}\frac{1}{p\left(x_{i}\right)},
\]
 and $\mu=\frac{N}{Z_{N}}$, as well as 

\[
\nu=\frac{1}{Z_{N}}\sum_{i=1}^{N}\frac{\left|\phi_{s}\left(x_{i}\right)\right|^{2}}{p\left(x\right)}.
\]
 The overlap between $\psi^{\left(q,p\right)}$, and the Coulomb ground
state $\psi_{\mathrm{Coulomb}}$, is then computed as 
\[
\braket{\psi_{\mathrm{Coulomb}}}{\psi^{\left(q,p\right)}}=\mathcal{N}\sum_{s}\beta_{s}\braket{\phi_{s}}{\psi^{\left(q,p\right)}},
\]
 where $\beta_{s}=\braket{\psi_{\mathrm{Coulomb}}}{\phi_{s}}$ is
obtained from exact diagonalization. The normalization $\mathcal{N}$
is chosen such that $\mathcal{N}^{2}\sum_{s}\left|\braket{\phi_{s}}{\psi^{\left(q,p\right)}}\right|^{2}=1$. 

It is possible to perform numerical comparisons for only a small number
of electrons, since the dimension of the LLL grows exponentially in
the number of electrons. For $N_{e}=3$, electrons the LLL has $10$
many particle states and for $N_{e}=4$ electrons, the LLL has $43$
states. These Hilbert spaces are still rather small, but the numerical
complexity comes about since the overlap $\braket{\psi^{\left(q,p\right)}}{\phi_{s}}$
has to be calculated for all the basis states $\phi_{s}$%
\footnote{In retrospect, a more effective algorithm would have been to compare
with the Coulomb energy eigenstates, as their expected overlap with
$\psi^{\left(q,p\right)}$ should fall off with energy. No such statement
can be made for the momentum basis states.%
}. It is the Monte Carlo sampling of all of these states that take
the majority of the time. As the number of electrons are increased,
the number $N$ of Monte Carlo coordinates needed, also increase.
This also affects the numerical complexity.

For $N_{e}=3$, taking $N=3\times10^{6}$ Monte Carlo points at $\tau=\i$,
the $(q,p)=(3,2)$ state has the best overlap with exact Coulomb,
$|\braket{\psi_{\mathrm{Coulomb}}}{\psi^{(3,2)}}|^{2}=0.9999(4\pm6)$
as compared to Laughlin, which has $|\braket{\psi_{\mathrm{Coulomb}}}{\psi_{\mathrm{Laughlin}}}|^{2}=0.9990(0\pm2)$.

For $N_{e}=4$, taking $N=3\times10^{7}$ Monte Carlo points, the
$(q,p)=(3,1)$ state matches Coulomb best, with $|\braket{\psi_{\mathrm{Coulomb}}}{\psi^{(3,1)}}|^{2}=0.9976(5\pm6)$
compared to $|\braket{\psi_{\mathrm{Coulomb}}}{\psi_{\mathrm{Laughlin}}}|^{2}=0.9792(8\pm3)$
for the Laughlin state. 

We see numerically that we can improve on the Laughlin state at $\nu=\frac{1}{q}$,
by considering $\psi^{(q,p)}$ with $p\neq0$. This result is in agreement
with Girvin and Jach\cite{Girvin_84} on the plane.

\subsection{How to Treat the Derivatives in Many-Particle States\label{sec:Derivatives}}

As seen in the previous section, we can write trial wave functions
for states in the hierarchy using conformal blocks. On the plane and
on the sphere, a trial state can be generated for any rational filling
fraction. The method of using conformal blocks and primary operators
for describing the electrons, usually results in derivatives, that
act on the higher level condensates. The simplest example of this
is the case of $\nu=\frac{2}{5}$, that has the electron operators
\[
V_{1}\left(w\right)=e^{\i\sqrt{3}\phi_{1}\left(w\right)}\qquad V_{2}\left(z\right)=\partial_{z}e^{\i\frac{2}{\sqrt{3}}\phi_{1}\left(z\right)+\i\sqrt{\frac{5}{3}}\phi_{2}\left(z\right)}.
\]
 We will here use different notation for the electrons described by
$V_{1}$ and $V_{2}$, to emphasise that not all electrons are treated
equally. Because of this asymmetry, the full many-body wave function
needs to be antisymmetrized at the very end of the calculation. When
calculating the trial wave functions in a planar geometry, the correlator
can be factorized such that the derivatives are outside of the correlator.
The trial wave functions are then calculated as
\[
\psi=\prod_{j}\partial_{z_{j}}\left\langle \mathcal{O}_{\mathrm{bg}}\prod_{i}V_{1}\left(w_{i}\right)\cdot\prod_{j}\hat{V}_{2}\left(z_{j}\right)\right\rangle .
\]
Here $\hat{V}_{2}\left(z\right)$ is the electron operator without
a derivative, such that $V_{2}\left(z\right)=\partial_{z}\hat{V}_{2}\left(z\right)$.
As always, a background charge is inserted to make the whole correlator
charge neutral. On the torus, things are more complicated, as the
correlator can not directly be factorized in a holomorphic and an
anti-holomorphic component. Also the treatment of the derivatives
is somewhat obscure, as these should now be acting within the full
correlator. The approach taken here, is to ignore the external derivatives
and insert them first at the very end of the calculation. The correlator
can now be calculated, and for $\nu=\frac{2}{5}$ the linear combination
of conformal blocks that fulfils the boundary conditions are 
\begin{eqnarray}
 &  & \psi_{s}^{\left(\frac{2}{5}\right)}\left(\left\{ z\right\} ,\left\{ w\right\} \right)=e^{-\frac{1}{2}\sum_{j}w_{j}^{2}}e^{-\frac{1}{2}\sum_{j}z_{j}^{2}}\mathcal{H}_{s}\left(Z^{\left(1\right)},Z^{\left(2\right)}\right)\times\qquad\qquad\qquad\nonumber \\
 &  & \times\prod_{i<j}\elliptic 1{\frac{z_{i}-z_{j}}{L_{x}}}{\tau}^{3}\prod_{i<j}\elliptic 1{\frac{w_{i}-w_{j}}{L_{x}}}{\tau}^{2}\prod_{i,j}\elliptic 1{\frac{w_{i}-z_{j}}{L_{x}}}{\tau}^{2}.\label{eq:nu=00003D2/5_trial_function}
\end{eqnarray}
 The centre of mass function $\mathcal{H}_{s}\left(Z^{\left(1\right)},Z^{\left(2\right)}\right)$
is in turn calculated as a combination of conformal blocks 
\[
\mathcal{H}_{s}\left(Z^{\left(1\right)},Z^{\left(2\right)}\right)=\sum_{l=1}^{3}\left(-1\right)^{tl}\mathcal{G}_{2l}^{\left(1\right)}\left(Z^{\left(1\right)},Z^{\left(2\right)}\right)\mathcal{G}_{5l+3s}\left(Z^{\left(2\right)}\right),
\]
where $\mathcal{G}^{\left(j\right)}$ essentially are $\vartheta$-functions,
while $Z^{\left(1\right)}=3\sum_{j}z_{j}+2\sum_{j}w_{j}$ and $Z^{\left(2\right)}=5\sum_{j}w_{j}$\cite{Hermanns_08}.
In the $\nu=\frac{2}{5}$ case, as in general for hierarchy states,
all electronic coordinates are not equivalent in the construction
and a final antisymetrization of $\left\{ z\right\} $ and $\left\{ w\right\} $
is needed.

By evaluating \eqref{eq:nu=00003D2/5_trial_function} we might think,
that the problem of constructing a torus trial wave function has been
solved; Just put back the derivatives and all will be well. Alas,
we are not that fortunate. Since $\left[\partial_{z},\translate{\tau L_{x}}\right]\neq0$
the derivatives change the boundary conditions of the $\psi^{\left(\frac{2}{5}\right)}$
state and therefore can not be used. But why not skip the use of the
derivatives altogether? The answer is, that if we skip the derivatives
completely, the antisymetrized wave functions vanish identically.
Some analogue of derivatives must exist to prevent the antisymmetrization
from killing the trial wave function.

There is\emph{ à priori} no method telling us what should replace
the derivatives, when on the torus. There are some constraints that
limit the possible alternatives; The wave function should, in the
planar $N_{s}\rightarrow\infty$ limit, reduce to the planar wave
functions; The wave function should transform nicely under modular
transformations.

An appealing alternative would be to act with the derivatives, and
then project to the LLL. Doing this for one particle, we find the
projection to be 
\begin{equation}
\plll\partial_{z}\psi\left(z\right)=\frac{1}{2\i}\sum_{l=1}^{N_{s}}a_{l}t_{1}^{l}\psi\left(z\right),\label{eq:P_lll_to_translations}
\end{equation}
 if $\psi\left(z\right)$ is a LLL wave function\cite{Fremling_13a}.
The coefficients $a_{l}$ are given as the discrete Fourier transform
of a piecewise Gaussian function $G_{s}$, defined as 
\begin{equation}
G_{s}=y_{s}-\frac{L_{y}}{2\sqrt{\pi}}\sum_{t}\int_{tL_{2}-y_{s}+\delta}^{tL_{2}+y_{s}-\delta}dy\, e^{-y^{2}}.\label{eq:Definition_G_s}
\end{equation}
 Here $\delta$ is a free parameter describing how the torus is parametrized.
The fact that $\delta$ exists in the final result \eqref{eq:Definition_G_s},
shows that something is pathological, as we do not wish that the final
result depends on a parametrization. For the moment, setting aside
this caveat about the proper choice of $\delta$, we at least obtain
a method that translates derivatives $\partial_{z}$, to a well defined
operator $\mathcal{D}=\frac{1}{2\i}\sum_{l=1}^{N_{s}}a_{l}t_{1}^{l}$
on the torus. Applying the recipe $\partial_{z}\rightarrow\mathcal{D}$
on the conformal block given by \eqref{eq:nu=00003D2/5_trial_function}
we get the many-body wave function
\begin{eqnarray*}
\Psi_{s}\left(\left\{ z\right\} ,\left\{ w\right\} \right) & = & \prod_{j}\mathcal{D}_{w_{j}}\psi_{s}^{\left(\frac{2}{5}\right)}\left(\left\{ z\right\} ,\left\{ w\right\} \right)\\
 & = & \prod_{j}\left(\frac{1}{2\i}\sum_{l=1}^{N_{s}}a_{l}t_{1,w_{j}}^{l}\right)\psi_{s}^{\left(\frac{2}{5}\right)}\left(\left\{ z\right\} ,\left\{ w\right\} \right).
\end{eqnarray*}
 Formally we have managed to obtain a LLL wave function, \emph{but},
it is still pathological. The problem is the many-body operator $D_{w}=\prod_{j}\mathcal{D}_{w_{j}}$
itself. It is straightforward to verify that $D_{w}$ does not commute
with any power of $T_{2}$. This non-commutativity is really disastrous,
since it means that $D$ changes the quantum numbers of $\psi_{s}^{\left(\frac{2}{5}\right)}$,
and takes us out of the desired five-fold subspace of trial wave functions.
Thus, we can \emph{not} use equation \eqref{eq:P_lll_to_translations},
even if we find a proper choice of $\delta$.

It turns out that the only parts of $D_{w}$, that will commute with
$T_{2}^{5}$, are the parts that can be written as $T_{1,w}^{l}$.
Let us just take a moment and clarify the notation. The many-body
translation operators acting on the $w$ and $z$ coordinates are
written $T_{j,w}=\prod_{k}t_{j,w_{k}}$ and $T_{j,z}=\prod_{k}t_{j,z_{k}}$,
such that $T_{j}=T_{j,w}T_{j,z}$ act on all coordinates. 

To summarize, when constructing states that preserves the $q$-fold
degeneracy, only certain terms of $D_{w}$ are allowed to be kept.
The maximally allowed sum of terms may be written as

\begin{equation}
\Psi_{s}=\sum_{l=1}^{N_{s}}\alpha_{l}T_{1,w}^{l}\psi_{s}\left(\left\{ z\right\} ,\left\{ w\right\} \right),\label{eq:nu=00003D2/5_T1}
\end{equation}
 where the parameters $\alpha_{l}$ are unspecified for the time being.
This anzats is exactly what was used by Hermanns \emph{et.al.} in
Ref. \cite{Hermanns_08}. In their work, they found that as $L_{x}\rightarrow0$,
the first term $\alpha_{1}$ becomes increasingly dominant when fitted
to the exact ground state for the Coulomb potential.

\begin{figure}
\begin{centering}
\begin{tabular}{cc}
\includegraphics[width=0.45\columnwidth]{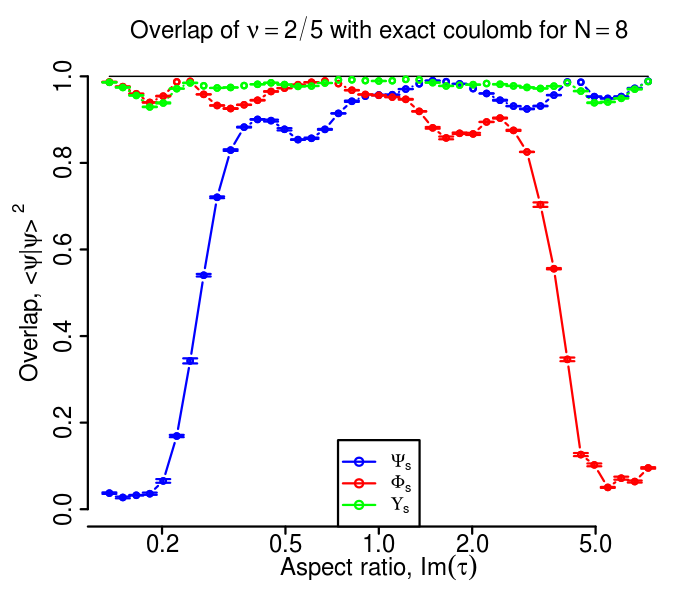} & 
\includegraphics[width=0.45\columnwidth]{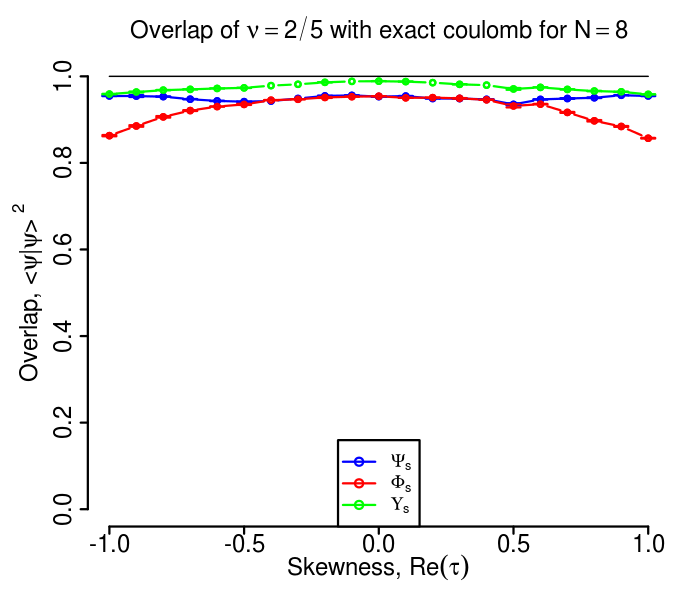}\tabularnewline
$a)$ & $b)$\tabularnewline
\end{tabular}
\par\end{centering}

\caption{Overlap between the exact coulomb ground and $\Psi_{s}$ (\textcolor{blue}{Blue}),
$\Phi_{s}$ (\textcolor{green}{Green}) and $\Upsilon_{s}$ (\textcolor{red}{Red}),
defined in equations \eqref{eq:nu=00003D2/5_T1}, \eqref{eq:nu=00003D2/5_T2}
and \eqref{eq:Final_nu=00003D2/5_anzats}. $\Psi_{s}$ and $\Phi_{s}$
have only $\alpha_{1}$ and $\beta_{1}$ non-zero. The number of electrons
are $N_{e}=8$. $a)$ Cross section of $\Re\left(\tau\right)=0$ for
$0.13<\Im\left(\tau\right)<7.4$. Notice that the overlap with $\Psi_{s}$
drastically vanishes as $\Im\left(\tau\right)\rightarrow0$. The mirrored
behaviour is seen for $\Phi_{s}$ as $\Im\left(\tau\right)\rightarrow\infty$.
The combination $\Upsilon_{s}$ is good for all values of $\Im\left(\tau\right)$.
\textbf{$b)$ }Cross section of $\Im\left(\tau\right)=1$ for $-1<\Re\left(\tau\right)<1$.
The combination $\Upsilon_{s}$ is still good even though non-trivial
phases enter through the coefficients $\alpha_{1}$ and $\beta_{1}$.\label{fig:T1_T2_overlap}}
\end{figure}

We now have a problem: We find that when $L_{x}$ is changed in the
opposite direction, such that $L_{y}\rightarrow0$, we see that \emph{no}
combination of $\alpha_{l}$ can give good overlap with the Coulomb
ground state. We can understand this result physically by considering
the torus geometry. When $L_{x}\rightarrow0$, the operator $t_{1,w}\approx1+\frac{L_{x}}{N_{s}}\partial_{x}$
approximates a derivative well, since the torus is thin in the $x$
direction. When $L_{y}\rightarrow0$, such that $L_{x}\rightarrow\infty$,
the torus is thin in the opposite direction. Now, the $t_{1,w}$ operator
does not resemble a derivative any more. We can remedy this by generalizing
the anzats \eqref{eq:nu=00003D2/5_T1}. We simply trade $T_{1}$ for
$T_{2}$. These two operators do \emph{not} commute, so $T_{2}$ will
change the momentum sector of the $\psi_{s}$ wave function. This
is easily accounted for by letting $\psi_{s}\rightarrow\psi_{s-k}$.
We thus get an alternative set of wave functions 

\begin{equation}
\Phi_{s}=\sum_{k=1}^{N_{s}}\beta_{k}T_{2,w}^{k}\psi_{s-k}\left(\left\{ z\right\} ,\left\{ w\right\} \right).\label{eq:nu=00003D2/5_T2}
\end{equation}
The numerical overlap with this function and the ground state of the
Coulomb potential is bad when $L_{x}\rightarrow0$, and good then
$L_{y}\rightarrow0$. This is the mirrored behaviour from $\Psi_{s}$,
as is seen in Figure \ref{fig:T1_T2_overlap}a. The general trial
wave function ansatz for the $\nu=\frac{2}{5}$ can thus be extended
to 
\begin{equation}
\Upsilon_{s}\left(\left\{ z\right\} \right)=\sum_{l=1}^{N_{s}}\left[\alpha_{l}T_{1,w}^{l}+\beta_{l}T_{2,w}^{l}T_{2}^{-l}\right]\psi_{s}\left(\left\{ z\right\} ,\left\{ w\right\} \right)\label{eq:nu=00003D2/5_full_anzats}
\end{equation}
 where we write $\psi_{s-k}=T_{2}^{-k}\psi_{s}$. How can we find
some guiding principle that can fix the values of $\alpha_{l}$ and
$\beta_{l}$? To find out, we need to study the modular behaviour
of the wave function \eqref{eq:nu=00003D2/5_full_anzats}. In doing
so, we will also realize that mixed terms, such as $T_{1,w}^{l}T_{2,w}^{k}$,
will also be needed in the anzats.

\subsection{The Requirement of Modular Covariance}

\begin{figure}
\begin{centering}
\begin{tabular}{cc}
\includegraphics[width=0.5\textwidth]{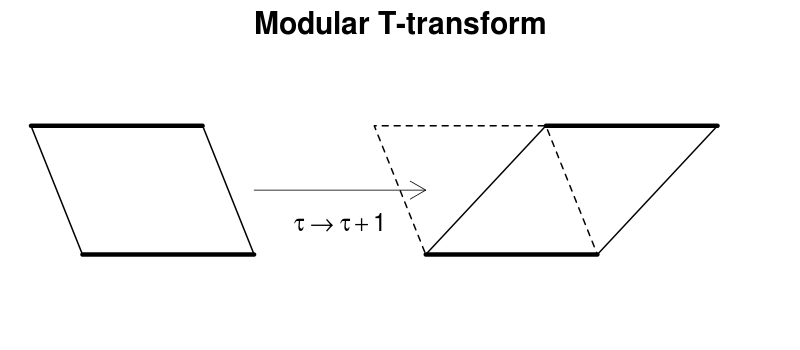} & 
\includegraphics[width=0.5\textwidth]{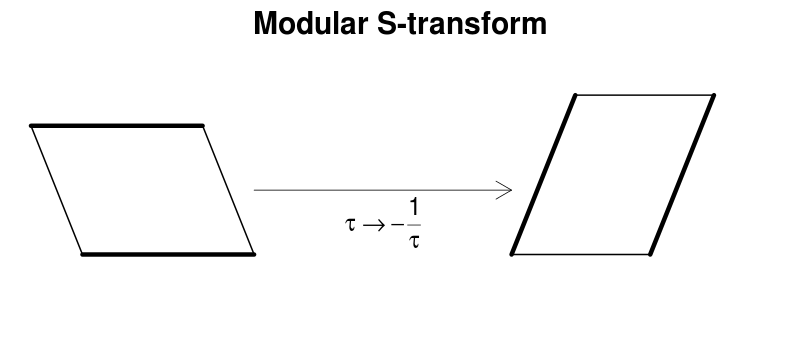}\tabularnewline
$a)$ & $b)$\tabularnewline
\end{tabular}
\par\end{centering}

\caption{$a)$ The geometric interpretation of a $\mathcal{T}$-transform,
$\tau\rightarrow\tau+1$. Both $L_{x}$and $L_{y}$ are unchanged
but the torus is tilted such that $L_{\Delta}\rightarrow L_{\Delta}+L_{x}$.
$b)$ The geometric interpretation of an $\mathcal{S}$-transform,
$\tau\rightarrow-\frac{1}{\tau}$. The torus is effectively rotated
such that $L_{x}\rightarrow\left|\tau\right|L_{x}$, $L_{y}\rightarrow\frac{1}{\left|\tau\right|}L_{y}$
and $L_{\Delta}\rightarrow-\frac{1}{\left|\tau\right|}L_{\Delta}$.\label{fig:S-T-transform}}
\end{figure}

Modular properties are important since they tell us about how $\Upsilon_{s}$
transforms under changes in $\tau$. The parameter $\tau$ encodes
information about the geometry of the torus, and thus the space the
electrons live on. Since $\tau$ measures the geometry, or more precisely
lets us know which points that are equivalent, there exists transformations
of $\tau$ that should not change the physics. One such transformation
is $\tau\rightarrow\tau+1$, also known as a $\mathcal{T}$-transformation.
Physically this just maps one point of the lattice onto the next point
so the geometry of the torus is effectively unchanged, ans can bee
seen in Figure \ref{fig:S-T-transform}a.

A second transformation we can perform, is to let $\tau\rightarrow-\frac{1}{\tau}$,
which is called an $\mathcal{S}$-transformation. The $\mathcal{S}$-transformation
is really a rotation of the torus, by effectively swapping the two
axes of the torus, as can bee seen in Figure \ref{fig:S-T-transform}b.
For a rectangular torus it is easily seen since $\tau=\i\frac{L_{y}}{L_{x}}$
and $-\frac{1}{\tau}=\i\frac{L_{x}}{L_{y}}$ differ by letting $L_{x}$
and $L_{y}$ trade places. From this simple analysis we can see that
when we let $\tau$ go to $-\frac{1}{\tau}$ in $\Upsilon_{s}$ we
will change $T_{1}$ into $T_{2}$. That $\mathcal{S}$ will transform
$T_{1}$ in $T_{2}$ and vice versa is good, since it will constrain
the parameters $\alpha_{l}$ and $\beta_{l}$ in \eqref{eq:nu=00003D2/5_full_anzats}.
We will use that the space of $\Upsilon_{s}$ under $\mathcal{S}$
should transform into itself such that 
\begin{equation}
\mathcal{S}\Upsilon_{s}=\sum_{s^{\prime}}\lambda_{ss^{\prime}}\Upsilon_{s^{\prime}}.\label{eq:S-transform}
\end{equation}
 The complete analysis that fixes the $\tau$ dependence of $\alpha$
and $\beta$ relies on an analysis of the modular transformation properties
of the conformal blocks that build $\Upsilon_{s}$. A detailed description
of that procedure can be found in Ref. \cite{Fremling_13b}, and here
we make an heuristic argument of the general behaviour of $\alpha\left(\tau\right)$
and $\beta\left(\tau\right)$. The numerical results in Figure \ref{fig:T1_T2_overlap}
implies that $\frac{\alpha\left(\tau\right)}{\beta\left(\tau\right)}\rightarrow0$
as $\tau\rightarrow0$, and that $\frac{\alpha\left(\tau\right)}{\beta\left(\tau\right)}\rightarrow\infty$
as $\tau\rightarrow\infty$. The parameters $\alpha$ and $\beta$
must thus depend on $\tau$, and have certain limiting behaviour.
From \eqref{eq:S-transform} we know that $\alpha$ and $\beta$ must
transform into each other under $\mathcal{S}$, such that $\Upsilon_{s}$
has the proper modular behaviour. The last piece of the puzzle comes
from a full analysis of how the quasi-particle operators should be
regularized in the toroidal geometry. When all of the above factors
are taken into account, the coefficients for $\alpha_{1}$ and $\beta_{1}$
are

\[
\alpha_{1}=\left(\frac{1}{\elliptic 1{\frac{1}{N_{s}}}{\tau}}\right)^{\frac{N_{e}}{2}}\qquad\qquad\beta_{1}=\left(\frac{e^{-\i\pi\tau\frac{1}{N_{s}^{2}}}}{\elliptic 1{\frac{\tau}{N_{s}}}{\tau}}\right)^{\frac{N_{e}}{2}}.
\]
 The first two terms in $\Upsilon_{s}$ are therefore 
\begin{equation}
\Upsilon_{s}=\mathcal{N}\left(\tau\right)\left[\left(\frac{1}{\elliptic 1{\frac{1}{N_{s}}}{\tau}}\right)^{\frac{N_{e}}{2}}T_{1,w}\psi_{s}\left(w,z\right)+\left(\frac{e^{-\i\pi\tau\frac{1}{N_{s}^{2}}}}{\elliptic 1{\frac{\tau}{N_{s}}}{\tau}}\right)^{\frac{N_{e}}{2}}T_{2,w}\psi_{s-1}\left(w,z\right)\right].\label{eq:Final_nu=00003D2/5_anzats}
\end{equation}
 In the above equation, the full dependence on $\tau$ is hidden in
the overall normalization $\mathcal{N}$.

We can see from Figure \ref{fig:T1_T2_overlap} that the above anzats
gives good overlap with exact Coulomb for all values of $\tau$, not
only for $\Re\left(\tau\right)=0$. This is a non-trivial result since
the phases going into $\alpha_{1}$ and $\beta_{1}$ are strongly
fluctuating for a general $\tau$. A similar analysis will give us
relations between the generic $\alpha_{j}$ and $\beta_{j}$. The
modular $\mathcal{S}$ transformation does however not shed any light
on the relative size of the different $\alpha_{j}$, so we need another
mechanism to fix these. Some numerical work suggest that $\alpha_{j}\propto a_{j}^{\frac{N_{e}}{2}}$,
where $a_{j}$ are the coefficients defined in \eqref{eq:P_lll_to_translations},
gives an improved overlap with exact coulomb compared to $\alpha_{j}=\delta_{j,1}$.
This is of course a natural guess since the terms in $\prod_{j}\mathcal{D}_{w_{j}}$
that commutes with $T_{2}^{q}$, have precisely $\alpha_{j}\propto a_{j}^{\frac{N_{e}}{2}}$
as coefficients.

The $\mathcal{T}$ -transform introduces further constraints on $\Upsilon_{s}$.
Under $\mathcal{T}$, $\tau\rightarrow\tau+1$, the different powers
of $T_{1}$ and $T_{2}$ will transform into each other, such that
$T_{2}^{n}\rightarrow T_{1}^{n}T_{2}^{n}$. This is easily seen, since
what was a translation in the $\tau$-direction, will now be a translation
in the $\left(\tau+1\right)$-direction. That $T_{2}$ transforms
into $T_{1}T_{2}$ means that we have to extend the anzats in equation
\eqref{eq:nu=00003D2/5_full_anzats} to include all combinations of
$T_{1}$ and $T_{2}$, since they can all be reached by the two modular
transformations%
\footnote{Actually the space of $T_{1}^{m}T_{2}^{n}$ terms split into several
disjoint but self-similar sets. Two elements $T_{1}^{m}T_{2}^{n}$
and $T_{1}^{m^{\prime}}T_{2}^{n^{\prime}}$  can only be connected
through a combinations of $\mathcal{S}$ and $\mathcal{T}$ if their
greatest common divisor are the same, $\gcd\left(\left|m\right|,\left|n\right|\right)=\gcd\left(\left|m^{\prime}\right|,\left|n^{\prime}\right|\right)$.%
} 

\[
\mathcal{S}:T_{1}^{m}T_{2}^{n}\propto T_{1}^{n}T_{2}^{-m}
\]
 and 
\[
\mathcal{T}:T_{1}^{m}T_{2}^{n}\propto T_{1}^{m+n}T_{2}^{n}.
\]
 The extend anzats for $\Upsilon_{s}$, that is covariant under both
$\mathcal{S}$ and $\mathcal{T}$, is written as 
\begin{equation}
\Upsilon_{s}=\mathcal{N}\left(\tau\right)\sum_{n,m=1}^{N_{s}}\frac{e^{\i\lambda_{n,m}}}{\ellipticgeneralized{\frac{n}{N_{s}}+\frac{1}{2}}{\frac{m}{N_{s}}+\frac{1}{2}}0{\tau}^{\frac{N_{e}}{2}}}T_{1,w}^{m}T_{2,w}^{n}\psi_{s-n}\left(\left\{ w\right\} ,\left\{ z\right\} \right),\label{eq:Final_nu=00003D2/5_anzats_all_terms}
\end{equation}
 where the phase $\lambda_{n,m}$ is fixed by modular covariance.
We do not know, at this time, if \eqref{eq:Final_nu=00003D2/5_anzats_all_terms}
is the unique solution that respects both $\mathcal{S}$ and $\mathcal{T}$
transformations. 

As second check that the state \eqref{eq:Final_nu=00003D2/5_anzats_all_terms}
is properly describing a Quantum Hall fluid, we may calculate the
viscosity of that state. In the next Section, this will be done.

\pagebreak{}

\chapter[Viscosity in FQHS]{Viscosity in Fractional Quantum Hall States}

As shown before, there are novel difficulties when going to the torus,
as compared to the plane. After all, the wave functions are more complicated
and there is no clear analogy of what the derivatives are. So why
should we bother at all with the torus? We are still interested, because
some things are comparatively easy to calculate to the torus, but
difficult in other geometries. One such thing is the antisymmetric
component of the viscosity tensor. We will soon return to precisely
what the antisymmetric viscosity is and how it is calculated.

Our story begins on the sphere. When establishing the filling fraction
of a quantum Hall state, the number of fluxes $N_{\Phi}$ is compared
to the number of electrons $N_{e}$. On the torus, $N_{e}$ is proportional
to $N_{\Phi}$, such that $N_{e}=N_{s}\nu$, but this is not true
on the sphere. Because of the curved surface, and that the electron
has a spin, the electrons will pick up a Berry phase as it moves over
the surface of the sphere. This Berry phase will show up as a shift
$\mathcal{S}$, in the relation between $N_{e}$ and $N_{\Phi}$,
such that $N_{e}=\nu\left(N_{\Phi}+\mathcal{S}\right)$. For the IQHE
$\mathcal{S}=1$, but for the FQHE $\mathcal{S}>1$. For the Laughlin
state at $\nu=\frac{1}{3}$, we have $\mathcal{S}=3$ whereas for
$\nu=\frac{2}{5}$, we have $\mathcal{S}=4$. As such, the shift contains
information about the average orbital spin of the electrons $\bar{s}$,
such that $\mathcal{S}=2\bar{s}$. As different quantum Hall states
will have different shifts, it can be used to distinguish these states
from each other.

At first glance it looks as if the shift is a purely geometrical effect
and has noting to to with the torus, but this is incorrect. The shift
is a topological characteristic of the quantum Hall system, and must
thus be observable on all geometries. On the torus, which is a flat
surface, the orbital spin does not manifest itself in the filling
fraction equation, but rather as a transport coefficient. This particular
coefficient is the antisymmetric component $\eta_{A}$ of the viscosity
tensor. The antisymmetric viscosity is a peculiar thing. Whereas the
symmetric viscosity component, the shear viscosity $\eta_{S}$, is
related to dissipation, and can be thought of as the \emph{thickness
}of a fluid, the antisymmetric component is related to dissipationless
response. Simply put, if a system with $\eta_{A}\neq0$ is put under
strain, it will start to twist.

The particular type of viscosity we seek to calculate, sometimes called
the Hall viscosity, is unique for two-dimensional systems. Avron \emph{et
al.} computed the Hall viscosity for filled Landau Levels\cite{Avron_95}
and found the viscosity to be \textbf{$\eta_{A}=\frac{B}{8V}$}. In
the case of a partially filled LL, the analysis is more involved,
but has been performed by Read \& Rezayi for the Laughlin state and
the Moore-Read state\cite{Read_09,Read_11}.

Read has demonstrated, that the mean orbital spin is related to the
antisymmetric viscosity of the Quantum Hall system\cite{Read_09}
as $\eta_{A}=\frac{1}{2}\bar{s}\bar{n}\hbar$. In this simple formula
$\bar{n}$ is the number density of electrons and $\bar{s}$ the mean
orbital spin of each electron. It is therefore important to calculate
the viscosity for the $\nu=\frac{2}{5}$ trial wave function, to make
sure that it is in the right topological phase.

\section{Viscosity in the $\nu=\frac{2}{5}$ State}

Considering the the $\nu=\frac{2}{5}$ state, given by \eqref{eq:Final_nu=00003D2/5_anzats_all_terms},
two things need to be established. First, the overlap with the exact
Coulomb ground state has to be high. This will be the first test that
the wave function \eqref{eq:Final_nu=00003D2/5_anzats_all_terms}
is reasonable. As seen in Figure \ref{fig:T1_T2_overlap}, depending
on whether $\Im\left(\tau\right)$ is small or large, there is either
good overlap with $\Psi_{s}$ or with $\Phi_{s}$ from equations \eqref{eq:nu=00003D2/5_T1}
and \eqref{eq:nu=00003D2/5_T2}. Using the combination of both $\Psi_{s}$
and $\Phi_{s}$, with the parameters obtained from modular covariance,
the combined wave function has good overlap with exact coulomb for
all values of $\tau.$

We now have two systems for which we may compute the viscosity of
the state $\nu=\frac{2}{5}$. The exact coulomb ground state, and
the Hierarchy state given by \eqref{eq:Final_nu=00003D2/5_anzats_all_terms}.
These systems should both agree with the predicted value of $\bar{s}$.
We wish to compute the viscosity both for the exact diagonalization
and the trial wave functions, because overlap is not the full story.%
\footnote{There are examples of wave functions, that have very good overlap,
they still have very different symmetries. A case in point is the
Gaffnian\cite{Simon_07}, that has good overlap with the exact Coulomb
ground state, but also has several pathological properties. One of
these properties is the existence of gappless excitations, such that
the Gaffnian does not represent a stable gapped topological phase
of matter.%
}

Numerically the viscosity is calculated by evaluating the Berry curvature
$\mathcal{F}$ at a specific $\tau=\tau_{x}+\i\tau$. We may compute
the mean Berry curvature, in a region $\Omega$ by integrating the
Berry connection around a closed loop following the contours of $\Omega$.
We follow the numerical procedure of Read \& Rezayi\cite{Read_11}.
The mean Berry curvature $\bar{\mathcal{F}}$ is obtained from the
Berry connection as $\bar{\mathcal{F}}=\frac{1}{A_{\Omega}}\oint_{\partial\Omega}A_{\mu}\left(\lambda\right)\, d\lambda_{\mu}$.
If the area of $\Omega$ is small enough, $\mathcal{F}$ is almost
constant, and the path $\partial\Omega$ may be discretized. As a
result $\bar{\mathcal{F}}$ can be evaluated as 
\begin{equation}
e^{\i A_{\Omega}\mathcal{\bar{F}}}=e^{\i\oint A_{\mu}\left(\lambda\right)\, d\lambda_{\mu}}\approx\prod_{j}\braket{\varphi_{j}}{\varphi_{j+1}},\label{eq:Berry-curvature}
\end{equation}
 where $\ket{\varphi_{j}}$ is the state at point $j$. The area of
$\Omega$ is calculated as 
\begin{equation}
A_{\Omega}=\int_{\Omega}\frac{d\tau_{x}\, d\tau_{y}}{\tau_{y}^{2}}=2\pi\left[\frac{\tau_{y,0}}{\sqrt{\tau_{y,0}^{2}-\rho_{0}^{2}}}-1\right],\label{eq:Modular-invariant-area}
\end{equation}
 where $\tau_{y,0}$ is the imaginary $\tau$ coordinate for the centre
of $\Omega$, and $\rho_{0}$ is the radius of $\Omega$. The mean
orbital spin $\bar{s}$ can now be calculated as 
\begin{eqnarray}
\bar{s} & = & \frac{1}{2}+2\frac{\Re\left(\bar{\mathcal{F}}\right)}{A_{\Omega}N_{e}}.\label{eq:sbar-equation}
\end{eqnarray}

The added constant $\frac{1}{2}$ is the intrinsic spin of the electrons.
In general $\bar{\mathcal{F}}$ has a non-zero imaginary part, if
computed through \eqref{eq:Berry-curvature}. That is why explicitly
the real part of $\bar{\mathcal{F}}$ should enter into \ref{eq:sbar-equation}.

\begin{figure}
\begin{centering}
\includegraphics[width=0.6\columnwidth]{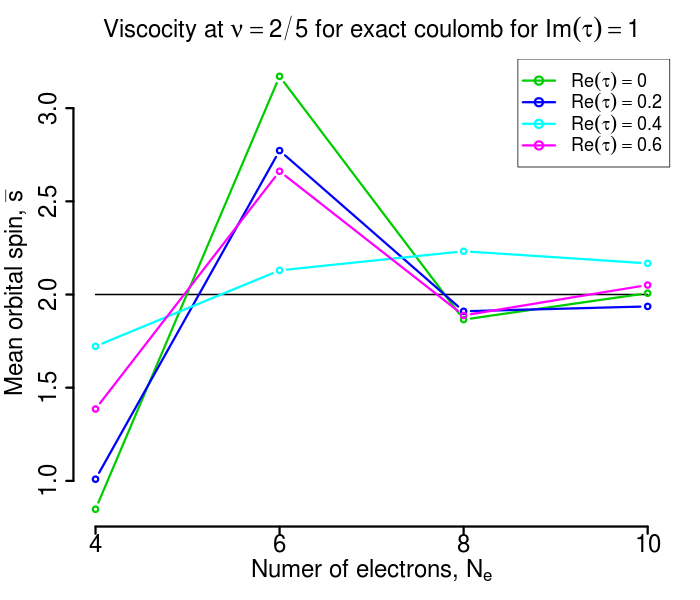}
\par\end{centering}

\caption{Viscosity, in units of the mean orbital spin $\bar{s}$, for the exact
coulomb ground, with for $N=4,\,6,\,8,\,10$ electrons. The torus
geometry is $\Im\left(\tau\right)=0$ and $\Re\left(\tau\right)=0,\,0.2,\,0.4,\,0.6$.
The different lines correspond to different skewness of the lattice.
The value of $\bar{s}$ depends on $\tau$, but it is likely a finite
size effect. This is seen since $\bar{s}$ converge on $\bar{s}=2$
as $N_{e}$ increases. \label{fig:Viscocity_colomb}}
\end{figure}

We calculate the viscosity of the exact coulomb ground state to be
$\bar{s}\approx2$ for $\nu=\frac{2}{5}$, even though the value of
$\bar{s}$ depends on $\tau$. This is depicted in Figure \ref{fig:Viscocity_colomb}.
We believe the $\tau$-dependence of $\bar{s}$ to be a finite-size
effect%
\footnote{It should be noted that for $\tau\rightarrow\i0$ and $\tau\rightarrow\i\infty$,
we expect $\bar{s}\rightarrow\frac{1}{2}$. In this limit, which is
the thin torus limit, all dynamics is frozen out and the problem becomes
one dimensional and electrostatic.%
}, as it becomes less pronounced for larger values of $N_{e}$. 

\begin{figure}
\begin{centering}
\includegraphics[width=0.6\columnwidth]{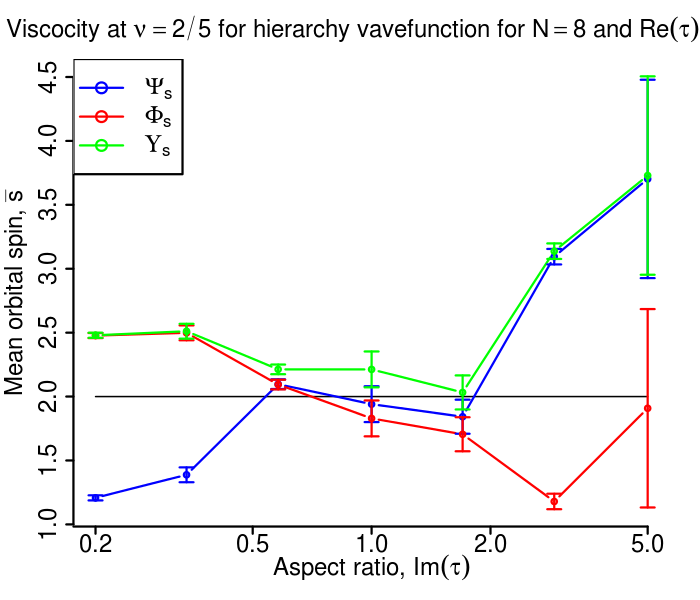}
\par\end{centering}

\caption{Viscosity, in units of the mean orbital spin $\bar{s}$, for the Hierarchy
wave functions $\Psi_{s}$ (\textcolor{blue}{Blue}), $\Phi_{s}$
(\textcolor{green}{Green}) and $\Upsilon_{s}$ (\textcolor{red}{Red})
defined in \eqref{eq:nu=00003D2/5_T1}, \eqref{eq:nu=00003D2/5_T2}
and \eqref{eq:Final_nu=00003D2/5_anzats}. The torus has the parameters
$\Re\left(\tau\right)=0$ and $0.2<\Im\left(\tau\right)<5$. In the
region $\Im\left(\tau\right)\approx1$ both $\Psi_{s}$ and $\Phi_{s}$
has viscosity near $\bar{s}=2$. For $\Im\left(\tau\right)\rightarrow0$
and $\Im\left(\tau\right)\rightarrow\infty$ the value $\bar{s}$
diverges, both from $\bar{s}=2$ and between $\Psi_{s}$ and $\Phi_{s}$.
It is not clear what is happening in these regimes, but is is likely
related to the torus becoming thin. It is clear that the different
weights in \eqref{eq:Final_nu=00003D2/5_anzats} are kicking in, as
$\Upsilon_{s}$ follows either $\Psi_{s}$ or $\Phi_{s}$ depending
on $\tau$. \label{fig:Viscocity_CFT}}
\end{figure}

We also compute the viscosity for the hierarchical wave function \eqref{eq:Final_nu=00003D2/5_anzats}.
There we find that the two components $\Psi_{s}$ and $\Phi_{s}$
have mutually diverging viscosity in the limit of a thin torus. This
is clearly seen in Figure \ref{fig:Viscocity_CFT}, where the coefficients
entering in \eqref{eq:Final_nu=00003D2/5_anzats} single out one viscosity
value over the other, as expected.

Numerically it is time consuming to evaluate the viscosity. For the
exact diagonalization we are as usual limited by the exponential growth
of the Hilbert space, meaning that it is not tractable to look at
systems larger than $N_{e}=12$. Also the number of steps that discretize
the path $\Omega$ should be on the order of $N=200$ steps, this
reduces the largest systems size to $N_{e}=10$.

For the Hierarchical states there is and added problem. Although we
do not need to perform an exact diagonalization, the overlap in \eqref{eq:Berry-curvature}
has to be evaluated using Monte Carlo methods. This introduces statistical
noise into the viscosity calculation.

To summarize: The Hierarchy states appear to have the mean orbital
spin $\bar{s}=2$ expected by studying the shift, although there are
still large numerical errors.

\pagebreak{}

\chapter{Summary and Outlook}

In this thesis we have mainly studied the behaviour of coherent states
on a torus. Further we have also touched upon the importance of modular
invariance when constructing trial wave functions on a torus. We have
looked explicitly on two alternative constructions for coherent states.
As an application of the coherent states, we have constructed a the
torus version of the modified Laughlin $\nu=\frac{1}{q}$ states,
as well as trial states for $\nu=\frac{2}{5}$, that is modular invariant.
In the case of the $\nu=\frac{5}{2}$ state, we have calculated the
viscosity, and found that it agrees with the predicted value.

The first approach was to project a Dirac $\delta$-functions on the
LLL. This produced a continuous set of wave functions that by necessity
where over-complete. These CCS fulfilled that same kind of resolution
of unity and self-reproducing kernel as the coherent states on the
plane, and they could be generated in the same way. It is fair so
say that the CCS are the torus analogue of the planar coherent states.

The second approach used Haldane and Rezayi's idea to place all zeros
at the same point. This generated a family of $N_{s}^{2}$ states
where $N_{s}$ is the number of fluxes in the system. These LCS wave
functions fulfilled relations, similar the resolution of unity and
self-reproducing kernel that the CCS possessed. The main difference
between the LCS and CCS, except for the number of existing states,
turns out to be their localization properties. The CCS approximately
minimize the uncertainty relations $\sigma_{x}\sigma_{y}$, for any
value of $\tau$, whereas the LCS form two distinct maxima at and
around $\Re\left(\tau\right)=\frac{1}{2}$.

Parts of this thesis deals with using the CCS and/or LCS basis, to
project functions that reside in higher Landau levels, down to the
lowest one. Interpreting a wave function in a basis of Coherent States,
is equivalent to projecting it to the LLL. This is crucial when the
trial wave functions contain anti-holomorphic components, such as
$\bar{z}$. On the plane, these anti-holomorphic coordinates can be
interpreted as a holomorphic derivative. On the torus, this interpretation
is not possible.

The issue of modular invariance ties in with the problems with derivatives.
We have shown that the derivative operator is ill-defined on the torus,
and has to be replaced by something else. A naive projection of $\partial_{z}$
on the LLL shows that $\plll\partial_{z}=\sum_{l}\alpha_{l}t_{1,z}^{l}$,
but since $\partial_{z}$ is ill-defined, the coefficients $\alpha_{l}$
can not be specified uniquely. We have shown that the many-body states
further restrict the values of $\alpha_{l}$, such that only terms
of the form $T_{1}=\prod_{l}t_{1,z_{l}}$, will preserve the $q$-fold
degeneracy of the trial wave functions. Because of modular covariance,
we have shown that there must also exist a term with $T_{2}=\prod_{l}t_{1,z_{l}}$.
The relative weights of $T_{1}$ and $T_{2}$ have been calculated
from modular covariance.

As a result of the covariance calculation, we have constructed trial
wave functions for the $\nu=\frac{2}{5}$ state. These wave functions
have good agreement with the Coulomb ground state, in the entire $\tau$
plane, already for only one $T_{1}$ and one $T_{2}$ term. This enabled
us to calculate the viscosity of the trial wave function, and find
that is coincides well with the values retrieved from exact diagonalization
of a Coulomb potential.

Future work will extend the $T_{1}$ and $T_{2}$ construction to
other filling fractions of the hierarchy. Using coherent states we
may also construct trial wave functions for filing fractions that
can not be reached through only particle condensation.

The antisymetrization of the $\nu=\frac{2}{5}$ state poses a numerical
problem, as the number of terms that needs to be evaluated grows as
$\left(\begin{array}{c}
2N_{e}\\
N_{e}
\end{array}\right)\sim2^{2N_{e}}$. Numerical methods that effectively perform Monte Carlo calculations,
using the $\nu=\frac{2}{5}$ trial wave function, as well as higher
hierarchy states, need to be developed.

\pagebreak{}

\appendix

\chapter{Jacobi Theta Functions and some Relations\label{sec:Jacobi-theta-functions}}

All LLL wave functions can be written as a Gaussian part and a holomorphic
function. On the torus, which is quasi two-dimensional an natural
set of functions to use are the Jacobi $\vartheta$-functions. In
this appendix we collect the main properties of these functions that
will be used throughout the main text. The generalized Jacobi $\vartheta$-function
is defined as 
\begin{equation}
\ellipticgeneralized abz{\tau}=\sum_{k=-\infty}^{\infty}e^{\i\pi\tau\left(k+a\right)^{2}}e^{\i2\pi\left(k+a\right)\left(z+b\right)}\label{eq:gen_theta_def}
\end{equation}
 where $\Im\left(\tau\right)>0$ for convergence. The zeros of \eqref{eq:gen_theta_def}
are located at 
\begin{equation}
z=\frac{1}{2}+m-b+\left(\frac{1}{2}+n-a\right)\tau.\label{eq:ge_theta_zeros}
\end{equation}
The $\vartheta$-function has two real parameters $a$ and $b$ that
fulfil 
\begin{equation}
\ellipticgeneralized{a+1}bz{\tau}=\ellipticgeneralized abz{\tau}\label{eq:gen_theta_a+1}
\end{equation}
 and 

\begin{equation}
\ellipticgeneralized a{b+c}z{\tau}=\ellipticgeneralized ab{z+c}{\tau}\label{eq:gen_theta_b+c}
\end{equation}

The two main periodic properties are

\begin{equation}
\ellipticgeneralized ab{z+n}{\tau}=e^{\i2\pi an}\ellipticgeneralized abz{\tau}\label{eq:gen_theta_z+n}
\end{equation}
 where $n\in\mathbb{Z}$ and 

\begin{equation}
\ellipticgeneralized ab{z+c\tau}{\tau}=e^{-\i2\pi c\left(z+b\right)}e^{-\i\pi\tau c^{2}}\ellipticgeneralized{a+c}bz{\tau}\label{eq:gen_theta_z+n_tau}
\end{equation}
 where $c\in\mathbb{R}$. Under transformations of the lattice parameter
$\tau$ the relations are
\begin{equation}
\ellipticgeneralized abz{\tau+n}=e^{-\i\pi a\left(1+a\right)n}\ellipticgeneralized a{an+\frac{n}{2}+b}z{\tau}\label{eq:gen_theta_z+tau+n}
\end{equation}
 where $n\in\mathbb{Z}$. Using the Poisson summation formula 

\[
\sum_{n\in\mathbb{Z}}e^{-\pi an^{2}+bn}=\frac{1}{\sqrt{a}}\sum_{k\in\mathbb{Z}}e^{\frac{\left(b+2\pi\imath k\right)^{2}}{4\pi a}}
\]
 we find that under inversion of the lattice parameter $\tau\rightarrow-\frac{1}{\tau}$,
the transformation is

\begin{eqnarray}
\ellipticgeneralized abz{-\frac{1}{\tau}} & = & \sqrt{-\i\tau}e^{\i\tau\pi z^{2}}e^{\i2\pi ba}\ellipticgeneralized b{-a}{\tau z}{\tau}\label{eq:gen_theta_taun_inverse}
\end{eqnarray}

There is a simple summation rule under Fourier sums 
\begin{equation}
\sum_{r=1}^{N}e^{\i\frac{2\pi}{N}rs}\ellipticgeneralized{a+\frac{r}{N}}bz{\tau}=e^{-\i2\pi as}\ellipticgeneralized{Na}{\frac{b+s}{N}}{\frac{z}{N}}{\frac{\tau}{N^{2}}}\label{eq:gen_theta_Fourier_sum}
\end{equation}

We can define four special cases of the parameters $a$ and $b$ that
that have symmetry properties under $z\rightarrow-z$. These functions
are 
\begin{eqnarray}
\elliptic 1z{\tau} & = & \ellipticgeneralized{\frac{1}{2}}{\frac{1}{2}}z{\tau}\label{eq:theta_1}\\
\elliptic 2z{\tau} & = & \ellipticgeneralized{\frac{1}{2}}0z{\tau}\label{eq:theta_2}\\
\elliptic 3z{\tau} & = & \ellipticgeneralized 00z{\tau}\label{eq:theta_3}\\
\elliptic 4z{\tau} & = & \ellipticgeneralized 0{\frac{1}{2}}z{\tau}\label{eq:theta_4}
\end{eqnarray}
where $\elliptic 1z{\tau}$ is odd and $\elliptic{2,3,4}z{\tau}$
are even.

\bibliographystyle{plain}
\bibliography{References}

\end{document}